\documentclass[pre, aps, showpacs, superscriptaddress, twocolumn, longbibliography]{revtex4-2}

\usepackage{hyperref}
\hypersetup{
     colorlinks=true,
     linkcolor=magenta,
     filecolor=blue,
     citecolor=blue,      
     urlcolor=cyan,
     }
\usepackage{color}
\usepackage[usenames,dvipsnames, table]{xcolor}
\usepackage{amsmath,amsthm,amssymb}
\usepackage{graphicx}
\usepackage{float}
\usepackage{bm}% bold math
\usepackage{mathrsfs}
\usepackage{multirow}
\usepackage{pbox}
\usepackage{verbatim}
\usepackage{braket}
\usepackage{mathtools}
\usepackage{bm}
\usepackage{mathtools}
\usepackage{tabstackengine}
\usepackage{enumerate}   
\usepackage{wasysym}
\usepackage{dsfont}
\usepackage{soul}
\usepackage{tabularx}
\usepackage{booktabs}
\usepackage{colortbl}

\stackMath
\DeclareMathOperator{\Tr}{Tr}

\newcommand{\er}[1]{Eq.~\eqref{#1}}

\newcommand{\beq}{\begin{equation}}
\newcommand{\eeq}{\end{equation}}

\DeclareMathOperator*{\argmin}{argmin}
\DeclareMathOperator*{\argmax}{argmax}

\begin{document}  

\title{Scalable simulation of non-equilibrium quantum dynamics via classically optimised unitary circuits}

\author{Luke Causer}
\affiliation{School of Physics and Astronomy, University of Nottingham, Nottingham, NG7 2RD, UK}
\affiliation{Centre for the Mathematics and Theoretical Physics of Quantum Non-Equilibrium Systems,
University of Nottingham, Nottingham, NG7 2RD, UK}
\author{Felix Jung}
\affiliation{Department of Physics, TFK, Technische Universität München, James-Franck-Straße 1, D-85748 Garching, Germany}
\author{Asimpunya Mitra}
\affiliation{Department of Physics, Indian Institute of Technology Kharagpur, Kharagpur 721302, India}
\affiliation{Department of Physics, University of Toronto, Toronto, Ontario M5S 1A7, Canada}
\author{Frank Pollmann}
\affiliation{Department of Physics, TFK, Technische Universität München, James-Franck-Straße 1, D-85748 Garching, Germany}
\affiliation{Munich Center for Quantum Science and Technology (MCQST), Schellingstr. 4, 80799 München, Germany}
\author{Adam Gammon-Smith}
\affiliation{School of Physics and Astronomy, University of Nottingham, Nottingham, NG7 2RD, UK}
\affiliation{Centre for the Mathematics and Theoretical Physics of Quantum Non-Equilibrium Systems,
University of Nottingham, Nottingham, NG7 2RD, UK}

\begin{abstract}
    The advent of near-term digital quantum computers could offer us an exciting opportunity to investigate quantum many-body phenomena beyond that of classical computing.
    To make the best use of the hardware available, it is paramount that we have methods that accurately simulate Hamiltonian dynamics for limited circuit depths.
    In this paper, we propose a method to classically optimise unitary brickwall circuits to approximate quantum time evolution operators.
    Our method is scalable in system size through the use of tensor networks.
    We demonstrate that, for various three-body Hamiltonians, our approach produces quantum circuits that can outperform Trotterization in both their accuracy and the quantum circuit depth needed to implement the dynamics, with the exact details being dependent on the Hamiltonian.
    We also explain how to choose an optimal time step that minimises the combined errors of the quantum device and the brickwall circuit approximation.    
\end{abstract}

\maketitle

\section{Introduction}
Quantum computers and quantum simulators provide the potential for a quantum advantage in complex computational problems \cite{Shor1997}, such as quantum chemistry \cite{AspuruGuzik2005, Kassal2011, Malley2016, Kandala2017,Cao2019,McArdle2020,Google2020,Bauer2020} and quantum machine learning \cite{Biamonte2017, Garcia2022, Zeguendry2023}.
It is expected that one of the earliest applications will be in the field of quantum many-body physics \cite{Feynman1982, Lloyd1996, Zalka1998, Preskill2018}.
Broadly speaking, the current methods to study quantum many-body systems fall into one of the three scenarios:
(i) exact numerics for small enough system sizes where the Hamiltonian can be diagonalised \cite{Zhang2010, Jung2020},
(ii) approximate numerics through low-rank approximations such as tensor networks for quantum states which have sufficiently small (area-law) entanglement \cite{Verstraete2006, Verstraete2006b, Hastings2007, Eisert2010, Dalzell2019},
or (iii) exact results for models which are simple enough to allow for a comprehensive analytical treatment \cite{Bertini2018, Skinner2019, Piroli2020, Buca2021,Zadnik2021,Zadnik2021b, Bertini2022, Fritzsch2023, Bertini2023}.
The limiting factor for (i) and (ii) is the curse of dimensionality; the Hilbert space grows exponentially in the number of degrees of freedom (e.g. the number of qubits).
Digital quantum computers, however, would allow us to faithfully simulate the dynamics of the system for large times through tunable quantum gates \cite{Lloyd1996, Tacchino2020}.
Such simulations could pave the way for a more thorough understanding of many important quantum phenomena such as thermalization and ergodicity \cite{DAlessio2016, Deutsch2018, Mori2018}.

Despite the prospects of quantum advantage \cite{Preskill2012, Arute2019}, modern-day quantum computers are still subject to strong noise, limiting the depths of the quantum circuits that can be simulated before the onset of decoherence. 
This provides us with a strong incentive to engineer optimal quantum circuits designed to achieve a specific task with minimal circuit depth.
Various approaches to achieve this have been proposed for the simulation of quantum many-body dynamics.
The most standard approach for simulating dynamics is to ``Trotterize'' the Hamiltonian \cite{Trotter1959, Suzuki1985}, which can be implemented on a quantum computer \cite{Lloyd1996, Zhukov2018, Lamm2018, Smith2019}.
However, to achieve large times with a desired accuracy, this can require a circuit depth that scales at least linearly in time \cite{Haah2018, Childs2019}.
One proposed method to overcome this complexity is to, at each time step, compress the quantum state as a unitary circuit of fixed depth \cite{Lin2021, Benedetti2021,Astrakhantsev2023}.
However, while this approach reduces the effects of noise from the quantum computer, it can also reduce the accuracy of the simulation.
Furthermore, it requires a costly optimisation task at each time step.

An alternative approach is to leverage classical computing to find optimal quantum circuits that can simulate quantum dynamics with a higher accuracy and smaller circuit depth when compared to Trotterization. 
On one hand, this could extend the times for which accurate simulations can be achieved.
On the other hand, unlike the approach described previously, would not require a costly optimisation task at each time step.
This approach was proven effective for the transverse field Ising model in Refs.~\cite{Mansuroglu2023, Kotil2023}, which optimised two-body brickwall circuits using gradient descent, and in Ref.~\cite{McKeever2023}, which made use of tensor networks for a restricted circuit ansatz.
A similar approach was taken in Ref.~\cite{Tepaske2023} for the Heisenberg model on a chain, and a ladder with next-nearest-neighbour interactions, which optimised over arbitrary two-qubit gates using automatic differentiation.

In this paper, we propose a tensor network (TN) approach to classically optimise brickwall (BW) unitary circuits.
Our method approximates the time propagator for Hamiltonian dynamics at small times as a matrix product operator (MPO) \cite{mcculloch2008, Chan2016, Hubig2017}.
This MPO is then used along with the {\em polar decomposition} \cite{Evenbly2009, Lin2021} to variationally optimise a unitary BW circuit on a classical computer by maximising their ``overlap''.
The advantage of using an MPO to approximate the time propagator is that it can reach a desired accuracy with resources which scale only linearly in system size \cite{Michalakis2012, Acoleyen2013, Colmenarez2020}.
The classically optimised circuit can thus be used as part of a larger calculation on a digital quantum computer to reach a total time larger than what is accessible on a classical computer.

Our paper is structured as follows.
In Sec.~\ref{sec: trotterization}, we give a brief introduction to simulating quantum dynamics with Trotterization.
The optimisation procedure is then explained in Sec.~\ref{sec: optimization}.
We demonstrate our approach in Sec.~\ref{sec: results} for various three-body Hamiltonians.
Using numerous metrics, we compare our results to Trotterization to demonstrate the practical advantage of classically optimised BW circuits.
We show that the optimised circuits outperform Trotterization in two ways: (i) they approximate the true time evolution operator with less error than Trotterization, and (ii) they can be implemented on digital quantum computers with less two-body entangling layers, thus reducing the error from decoherence.
We also benchmark convergence for our method.
In Sec.~\ref{sec: ibm}, we explore the trade-off between optimisation errors and decoherence errors from the device. We use these considerations to explain how to estimate the optimal time step which minimises the compounded error.
We give our conclusions in Sec.~\ref{sec: conclusions}.

\section{Trotterization} \label{sec: trotterization}
We consider locally interacting one-dimensional Hamiltonians, $\hat{H}$.
Although our optimisation method will be general, we will focus on local three-body interactions, 
\beq
    \hat{H} = \sum_{j=1}^{N} \hat{O}_{j}^{(1)} + \sum_{j=1}^{N-1} \hat{O}_{j, \, j+1}^{(2)} + \sum_{j=1}^{N-2} \hat{O}_{j, \, j+1, \, j+2}^{(3)},
    \label{H}
\eeq
where $\hat{O}_{j_{1}, \dots, j_{r}}^{(r)}$ describes some local operator which acts on lattice sites $j_{1}, \dots, j_{r}$.
We assume each lattice site has the same finite physical dimension $d$.
While, in principle, everything in this paper is applicable to arbitrary $d$, we will focus on the case of $d = 2$ (e.g. qubits).
One area of interest is quench dynamics: starting from some out-of-equilibrium initial state, $\ket{\psi_{0}}$, we study the dynamics $\ket{\psi_{t}} = U(t)\ket{\psi_{0}}$, where $U(t) = e^{-it\hat{H}}$ is the time evolution operator.
Calculating $\ket{\psi_{t}}$ for arbitrary $t$ can often be difficult due to the complexity of $U(t)$, which can be exponential in system size.

This difficulty could be overcome by simulating the Hamiltonian dynamics on a quantum computer.
We first use the product formula to slice the exponential into $n$ parts, $U(t) = U(\Delta t)^n$ for $\Delta t = t / n$ \cite{Lloyd1996}.
The unitary for the smaller time step $\Delta t$ can then be approximated by some other unitary ansatz that can be efficiently implemented on the quantum device, $V(\Delta t)$, such as the Trotter-Suzuki decomposition (see below for details) \cite{Trotter1959,Suzuki1985}.
The error of this unitary is 
\beq
    ||U(\Delta t) - V(\Delta t)|| \leq \mathcal{O}(\Delta t^{k+1}),
\eeq
where $k \geq 1$ is the {\em order} of the approximation and $|| \cdot ||$ is the operator norm.
It can then be shown that the total simulation error goes as \cite{Nielsen2011, Zhuk2023}
\beq
    || U(t) - V(\Delta t)^n || \leq n ||U(\Delta t) - V(\Delta t)||  = \mathcal{O}(t \Delta t^{k}).
\eeq
While only approximate, this approach offers the advantage that the error is well controlled through the choice of time step, $\Delta t$.

\begin{figure}[t]
    \centering
    \includegraphics[width=0.9\linewidth]{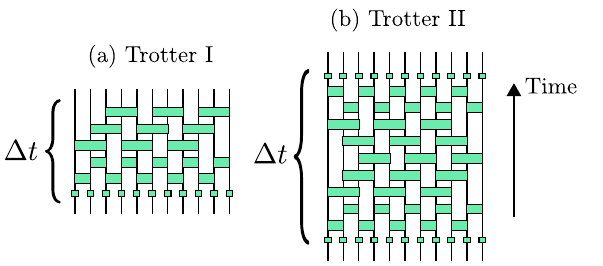}
    \caption{\textbf{Simulating time evolution with Trotterization.}
    The time evolution operator $U(\Delta t) = e^{-i\Delta t\hat{H}}$ can be approximated using the Trotter-Suzuki decomposition, as illustrated for local three-body interactions, see \er{H}, and $N = 11$ qubits.
    The lines indicate the direction of time for each qubit, and the green rectangles are local unitary operators.
    (a) The first order Trotter-Suzuki decomposition first evolves the system by all the single-body unitaries, followed by commuting two-body unitaries and finally three-body unitaries.
    Unitary operators which act on different qubits can be applied in parallel.
    Each unitary is the exponential of operators in \er{H} with time $\Delta t$.
    (b) The second order Trotter-Suzuki decomposition is done symmetrically in time: 
    all layers of unitaries are the exponential of operators with time $\Delta t / 2$, apart from the middle layer, which is done with time $\Delta t$.
    }
    \label{fig: trotterization}
\end{figure}

\subsection{Trotter-Suzuki decomposition} \label{sec: trotter}
A standard approach for approximating $U(\Delta t)$ is to decompose it into a product of {\em local} unitary operators, $U^{(r)}_{j_{1}, \dots, j_{r}}(\Delta t) = e^{-i\Delta t \hat{O}^{(r)}_{j_{1}, \dots, j_{r}}}$.
This is achieved through the Trotter-Suzuki decomposition \cite{Trotter1959, Suzuki1985}: for two operators $\hat{A}$ and $\hat{B}$, the first order Trotter-Suzuki decomposition allows us to write 
\beq
    e^{-i\Delta t(\hat{A} + \hat{B})} = e^{-i\Delta t\hat{B}}e^{-i\Delta t\hat{A}} + \mathcal{O}(\Delta t^{2}).
\eeq
In essence, this allows us to achieve the evolution by first evolving by $\hat{A}$, then followed by $\hat{B}$.
This has an error of order $\mathcal{O}(t^{2})$.
This is easily improved on through the second order Trotter-Suzuki decomposition, 
\beq
    e^{-i\Delta t(\hat{A} + \hat{B})} = e^{-i\Delta t\hat{A}/2}e^{-i\Delta t\hat{B}}e^{-i\Delta t\hat{A}/2} + \mathcal{O}(\Delta t^{3}).
\eeq
Like the first order decomposition, this evolves the state in a Floquet manner.
It first does a partial evolution over $\hat{A}$, followed by a full evolution over $\hat{B}$, and then a partial evolution over $\hat{A}$.
While this can be generalised to higher orders with more complex sequences of evolutions, here we will only consider up to the second order.

The Trotter-Suzuki decomposition allows for a way to approximately decompose $U(\Delta t)$ into a sequential application of local unitary operators.
For the Hamiltonian \er{H}, one must first split the Hamiltonian into various components which are each a sum over commuting operators.
An easy way to achieve this is to first write the Hamiltonian as $\hat{H} = \hat{H}^{(1)} + \hat{H}^{(2)} + \hat{H}^{(3)}$, with each the sum over all operators which act locally on $(r)$ qubits.
The first term, $\hat{H}^{(1)}$, is already a sum over commuting operators.
The second and third terms need to be split further,
\beq
    \hat{H}^{(2)} = \sum_{j} \hat{O}_{2j-1, 2j}^{(2)} + \sum_{j} \hat{O}_{2j, 2j+1}^{(2)} := \hat{H}^{(2)}_{1} + \hat{H}^{(2)}_{2},
\eeq
\begin{multline}
    \hat{H}^{(3)} = \sum_{j} \hat{O}_{3j-2, 3j-1, 3j}^{(3)} + \sum_{j} \hat{O}_{3j-1, 3j, 3j+1}^{(3)} \\
    + \sum_{j} \hat{O}_{3j, 3j+1, 3j+2}^{(3)} := \hat{H}^{(3)}_{1} + \hat{H}^{(3)}_{2} + \hat{H}^{(3)}_{3}.
\end{multline}
The Hamiltonian is now composed of six individual sums.
The Trotter-Suzuki decompositions can be applied five times in total to obtain a product over local unitary gates.
Since each sum contains commuting operators, this can be exactly split into the product of unitaries, e.g., $e^{-\Delta t\hat{H}^{(1)}} = \prod_{j} U^{(1)}_{j}(\Delta t)$.

Writing the full product of unitaries is rather cumbersome.
As such, it is convenient to write the full evolution diagrammatically, as shown in Figs.~\ref{fig: trotterization}(a, b) for the first and second order decompositions respectively. 
The lines show the evolution of a particular qubit (which are arranged in spatial order), the green rectangles show the action of a unitary gate on the given qubits.
For the first order decomposition, we first evolve by all single-body unitaries in parallel.
The next two layers demonstrate the application of two-body unitaries.
Notice that unitary gates that commute (act on different lattice sites) are applied in parallel.
Finally, we evolve over the three-body unitaries with three layers of gates.
For the second order decomposition, we evolve in a symmetric way: each gate is evolved for a time $\Delta t/2$, except for the middle layer of gates, which are evolved for the full time $\Delta t$.

\section{Classical optimisation of unitary circuits} \label{sec: optimization}
Trotterization provides a systematic way to approximate the time-evolution operator for a many-body quantum system by decomposing it into the product of local unitary operators. 
One drawback to this approach is that it still relies on many-body unitaries which might not be simple to implement on current quantum devices.
For example, the Trotter-Suzuki decomposition for the three-body Hamiltonian \er{H} would require us to implement three-body unitaries.
However, the native gates on popular platforms commonly only include two-body entangling gates, such as controlled-NOT (CNOT) gates.
While these provide a foundation for a universal set of quantum gates (i.e. any many-body unitary can be implemented to arbitrary accuracy), there is no promise that the three-body unitaries can be well implemented using a shallow quantum circuit.

\begin{figure}[t]
    \centering
    \includegraphics[width=0.9\linewidth]{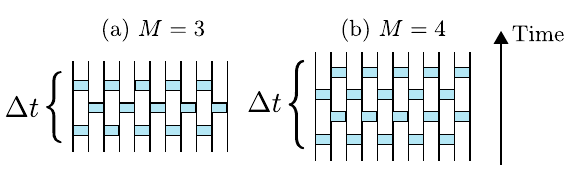}
    \caption{\textbf{Unitary brickwall circuits.}
    A unitary BW circuit can be used as a variational ansatz for the time-evolution operator $U(\Delta t)$, as shown for
     $N = 11$ qubits and a circuit depth of (a) $M=3$ and (b) $M = 4$. The lines indicate the direction of time for the circuit, and each rectangle is a two-qubit unitary matrix.
    }
    \label{fig: brickwall}
\end{figure}

In this section, we explain how one can optimise a BW unitary circuit with two-site gates and depth $M$ to approximate $U(\Delta t)$ (for small $\Delta t$) in a way that is scalable with system size $N$.
They are arranged compactly such that each layer has a unitary gate acting on each lattice site (except for lattice sites at the edges), see Figs.~\ref{fig: brickwall}(a, b) for visualisations of circuit depths $M = 3, 4$.
Note that, while there is no guarantee that a BW circuit with a fixed depth $M$ can well approximate $U(\Delta t)$, the operator entanglement of $U(\Delta t)$ for any local Hamiltonian is bounded by Lieb-Robinson bounds \cite{Lieb1972} and thus it is reasonable to expect that for sufficiently small times, $U(\Delta t)$ can be well-approximated by a BW circuit with a small depth \cite{Bertini2020}.
Additionally, unlike the Trotter circuits, there is no direct way to determine an optimal BW circuit for general $\hat{H}$.
However, they have the distinct advantage that they can be easily implemented on current quantum computers with a number of layers of two-body entangling gates that is constant in $M$, see App.~\ref{app: decompositions}.

\begin{figure}[t]
    \centering
    \includegraphics[width=0.5\linewidth]{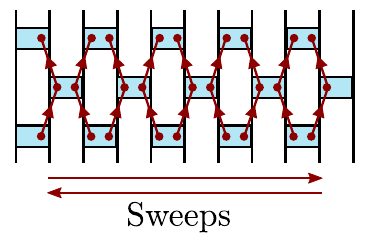}
    \caption{\textbf{Sweeping order.}
    We sweep back-and-forth through the lattice sites $j = 2$ to $N - 1$. 
    At each lattice site, we sequentially optimise each unitary gate which acts on site $j$ in the order shown in the diagram.
    }
    \label{fig: sweeps}
\end{figure}

We will optimise the BW unitary circuits to maximise the ``overlap'' with the true time evolution operator $U(\Delta t)$, which we approximate using an MPO.
Our method is reminiscent of standard variational matrix-product state methods and is not limited by system size (for example, Refs.~\cite{White1992, Rommer1997, Schollwock2011}): we find an optimal update for a single gate in the BW circuit, keeping all other gates fixed.
This is done for each gate in the circuit, which we sweep through multiple times until we find convergence.
While there is freedom in deciding the order to update the gates, the strategy we found to work best is given in Fig.~\ref{fig: sweeps}.
We sweep back and forth through the lattice sites $2 \leq j \leq N-1$.
For each lattice site, we update every gate which acts on the lattice site.
Every sweep from left-to-right (or right-to-left) will update each unitary gate twice, apart from those on the edges, which are only updated once.

\subsection{MPO approximation of $U(\Delta t)$}

\begin{figure}[t]
    \centering
    \includegraphics[width=\linewidth]{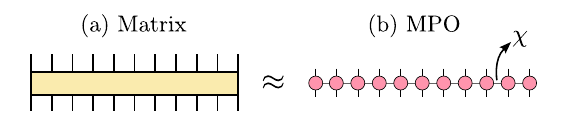}
    \caption{\textbf{Matrix product operators.}
    (a) The time propagator $U(\Delta t) = e^{-i\Delta t\hat{H}}$ can be represented as a full-rank matrix.
    (b) For small times, $U(\Delta t)$ can be well approximated with an MPO, where the pink circles represent rank-4 tensors.
    The virtual bond dimension, $\chi \ll d^{N} $, can be used to control the accuracy of the approximation.
    }
    \label{fig: mpo}
\end{figure}

In general, it is possible to write $U(\Delta t)$ as a $(d^{N}, d^{N})$-matrix.
This is illustrated diagrammatically as a TN in Fig.~\ref{fig: mpo}(a), where each leg corresponds to a physical dimension $d$.
However, this quickly becomes intractable to calculate due to its exponential cost with $N$.
One strategy to overcome this cost is to instead approximate $U(\Delta t)$ with an MPO.
This is a decomposition into a TN with a rank-4 tensor for each lattice site,
\begin{multline}
    U_{\rm MPO} = \sum_{n_{1}\cdots n_{N}} \sum_{m_{1}\cdots m_{N}} T_{n_{1}\cdots n_{N}}^{m_{1}\cdots m_{N}} \\ \ket{m_{1}\cdots m_{N}} \bra{n_{1} \cdots n_{N}},
\end{multline}
with the matrix coefficients determined by a matrix product,
\beq
    T_{n_{1}\cdots n_{N}}^{m_{1}\cdots m_{N}} = \Tr\left[ M^{(1)}_{n_{1}m_{1}} \cdots M^{(N)}_{n_{N}m_{N}} \right].
\eeq
Each $M^{(j)}_{n_{j}m_{j}}$ is a $(\chi, \chi)$-matrix, where $\chi$ is the virtual {\em bond dimension} which controls the accuracy of the approximation. 
When the labels $n_{j}$ and $m_{j}$ are unspecified, $M^{(j)}_{n_{j}m_{j}}$ is a rank-4 tensor with two physical dimensions and two bond dimensions which connect the tensor to the tensors of neighbouring lattice sites, see Fig.~\ref{fig: mpo}(b).
Each circle represents a tensor, the open legs the physical dimensions, and the connected legs the bond dimensions.

The MPO is described by $\mathcal{O}(N\chi^{2}d^{2})$ parameters, which is compared to the $\mathcal{O}(d^{2N})$ parameters of the full-rank matrix.
The caveat of this is that the MPO requires a bond dimension $\chi \sim \mathcal{O}(d^{N})$ to attain full accuracy for arbitrary $U(\Delta t)$.
Nevertheless, for small times, $U(\Delta t)$ can be approximated to some accuracy with a bond dimension which is independent of system size, $\chi \sim \mathcal{O}(d^{\gamma \Delta t})$ \cite{Michalakis2012, Acoleyen2013, Colmenarez2020}.
This fact will be used throughout the paper.

Our method will assume that we have access to a high accuracy MPO approximation of $U(\Delta t)$.
For system sizes $N\leq 12$, we construct the MPO by calculating $U(\Delta t)$ as a full-rank matrix to numerical precision, and then decompose it as an MPO using singular value decompositions (SVDs), see Ref.~\cite{Schollwock2011} for details.
For $N \geq 12$, we construct the approximation by evolving an MPO using Trotterization \cite{Verstraete2004}, although we note there are other methods which might be more effective, e.g., Refs.~\cite{Haegeman2011, Zaletel2015, Vandamme2023}.

We start with the identity matrix $U_{\rm MPO}(\tau = 0) = \hat{\mathds{1}}$.
To ensure the MPO approximation is accurate
\footnote{By this, we mean that the MPO has an accuracy much greater than that which can be achieved by the BW ansatz we are optimising.},
we evolve it using one-hundred second order Trotter steps with time step $\Delta \tau = \Delta t  / 100$, which we denote by $U_{\rm Trotter}(\Delta \tau)$.
At each evolution step, the time evolution operator is approximated by 
\beq
    U_{\rm MPO}(\tau + \Delta \tau) \approx U_{\rm Trotter}(\Delta \tau) \, U_{\rm MPO}(\tau).
\eeq
The application of each gate will give a new MPO with a bond dimension $\chi_{\rm exact} = r\chi$, where $r>1$ is some integer which depends on the circuit.
Thus, the bond dimension of the resulting MPO will grow exponentially in the number of applications of the Trotterized circuit.
To prevent this, at each time step, we approximate the MPO with another MPO with a smaller bond dimension $\chi'$ such that $\chi \leq \chi' \leq \chi_{\rm exact}$.
In practice, this is achieved by compressing the MPO using SVDs after each application of unitary gate \cite{Schollwock2011}.
We allow the bond dimension to grow dynamically, with its value determined through the SVD.
We use an SVD truncation error of $\epsilon = 10^{-16}$ to ensure that the truncation error is minimal
\footnote{See Ref.~\cite{Schollwock2011} for a definition of the truncation error.}.
Notice that this set-up allows for each individual bond dimension between lattice sites to take its own value.
Assuming a maximal bond dimension $\chi$, the cost of constructing $U(\Delta t)$ as an MPO is $\mathcal{O}(\Delta t \Delta \tau^{-1} \chi^{3} d^{8})$.

\subsection{The cost function}

\begin{figure}[t]
    \centering
    \includegraphics[width=\linewidth]{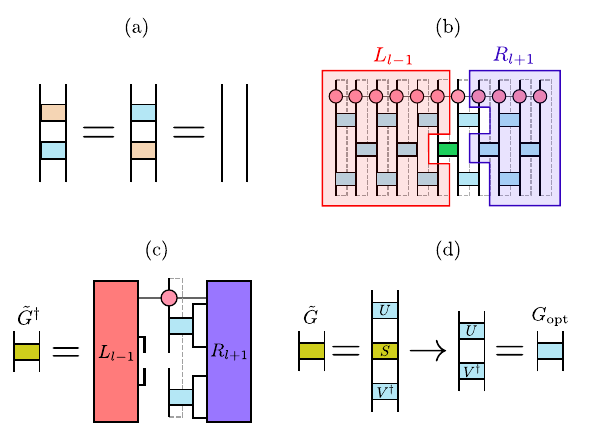}
    \caption{\textbf{Updating unitary gates.}
    The procedure for updating a gate $G$ illustrated for a circuit with depth $M = 3$.
    The target gate is at site $l = 7$ and depth $m = 2$.
    The optimal update for the gate is determined through \er{minimization}.
    (a) A unitary gate (blue rectangle), multiplied with its Hermitian adjoint (orange rectangle), gives the identity matrix.
    (b) The TN representation of $\Tr[U_{\rm MPO}^{\dagger}(\Delta t) U_{\rm BW}(\Delta t)]$.
    The green rectangle is the target gate.
    The tensors and gates which act on lattice sites $j < l$ (but not on $j = l$) can be contracted to give tensor $L_{l-1}$.
    Similarly, the tensors and gates which act on lattice sites $j > l$ (but not on $j = l$) can be grouped to give the tensor $R_{l+1}$.
    (c) The tensor $\tilde{G}^{\dagger}$ can be found by partially contracting over the TN in (c), excluding the gate $G$.
    (d) The choice of {\em unitary} gate which maximises ${\rm Re}(\Tr[U_{\rm MPO}^{\dagger}(\Delta t) U_{\rm BW}(\Delta t)])$ can be found through the polar decomposition. 
    In practice, we perform an SVD on the tensor $\tilde{G} = USV^{\dagger}$, and discard the matrix of singular values $S$.
    Multiplying $U$ and $V^{\dagger}$ gives the optimal unitary gate $G_{\rm opt} = UV^{\dagger}$.
    }
    \label{fig: update}
\end{figure}

Suppose we want to update the unitary gate at depth $m \leq M$ which acts on a lattice site $l$, which for simplicity we call $G$.
The objective is to choose the gate $G$ while keeping all other gates fixed such that the BW circuit best approximates the MPO.
To this end, we use the squared Frobenius norm
\begin{multline}
    \mathcal{F}^{(2)}( U_{\rm BW}(\Delta t)) = ||U_{\rm MPO}(\Delta t) - U_{\rm BW}(\Delta t)||^{2}_{F}
    \\
     = \Tr\left[U_{\rm MPO}^{\dagger}(\Delta t) U_{\rm MPO}(\Delta t)\right] + \Tr\left[U_{\rm BW}^{\dagger}(\Delta t) U_{\rm BW}(\Delta t)\right] 
     \\ - 2{\rm Re}\left(\Tr\left[U_{\rm MPO}^{\dagger}(\Delta t) U_{\rm BW}(\Delta t)\right]\right)
    \label{cost}
\end{multline}
as a cost function.
To find the optimal choice for the gate, $G_{\rm opt}$, one needs to minimise \er{cost} with respect to $G$, 
\beq
    G_{\rm opt} = \argmin_{G \in \mathcal{U}(d^{2})} \left\{ \mathcal{F}^{(2)}( U_{\rm BW}(\Delta t)) \right\},
    \label{minimization}
\eeq
where $\mathcal{U}(d^{2})$ denotes the group of unitary matrices with dimensions $(d^2, d^2)$.
The first term in \er{cost} depends only on the MPO, which has no variational parameters, and is thus constant.
The second term is the trace of the product of the BW circuit with its Hermitian adjoint.
Since all gates are unitary, the product of any gate with its Hermitian adjoint is the identity matrix, as shown in Fig.~\ref{fig: update}(a).
It follows that this term is constant, $\Tr[U_{\rm BW}^{\dagger}(\Delta t) U_{\rm BW}(\Delta t)] = 2^{N}$.
Note that this is true for {\em all} choices of unitary gates, $G$.
The final term is the trace of the BW circuit with the Hermitian adjoint of the MPO.
This is shown in Fig.~\ref{fig: update}(b).
The gate which is being optimised is coloured green.
Since the previous two terms are both constants, it follows that minimising the squared Frobenius norm is equivalent to
\beq
    G_{\rm opt} = \argmax_{G \in \mathcal{U}(d^2)} \left\{ {\rm Re}\left(\Tr\left[U_{\rm MPO}^{\dagger}(\Delta t) U_{\rm BW}(\Delta t)\right]\right) \right\}.
    \label{maximization}
\eeq

\subsection{Updating the gates}

We now explain how to find the unitary gate, $G_{\rm opt}$, which maximises ${\rm Re}(\Tr[U_{\rm MPO}^{\dagger}(\Delta t) U_{\rm BW}(\Delta t)])$.
We first contract all tensors and gates in $\Tr[U_{\rm MPO}^{\dagger}(\Delta t) U_{\rm BW}(\Delta t)]$ except for the gate $G$.
It is convenient to group together all tensors and gates which act on lattice sites $j < l$ (but not those which act on $j = l$) to give the tensor $L_{l-1}$.
Similarly, we group together all tensors and gates which act on lattice sites $j > l$ (but not those which act on $j = l$) to give the tensor $R_{l+1}$.
For more details, see Sec.~\ref{sec: environment}.
The gives the TN shown in Fig.~\ref{fig: update}(c), which can be fully contracted to give the tensor $\tilde{G}^{\dagger}$.
The exact contraction will differ depending on $l$, $m$ and $M$.
The cost of contracting the TN in Fig.~\ref{fig: update}(c) is at most $\mathcal{O}(\chi M d^{4 + 2\lceil M/2 \rceil} + \chi^{2} d^{2 + 2\lceil M/2 \rceil})$.
Note that, while the cost of contracting the network is exponential in the circuit depth $M$, in practice one wants to find a good approximation with as minimal depth as possible.

The unitary gate $G_{\rm opt}$ which maximises ${\rm Re}(\Tr[U_{\rm MPO}^{\dagger}(\Delta t) U_{\rm BW}(\Delta t)])$ can then be found through the {\em polar decomposition} of $\tilde{G}$ \cite{Evenbly2009, Lin2021}.
In practice, this is found by performing an SVD on $\tilde{G} = USV^{\dagger}$, where $U$ and $V^{\dagger}$ are unitary matrices, and $S$ is a diagonal matrix of singular values.
The polar decomposition can then be retrieved by replacing $S$ with identity, i.e., $G_{\rm opt} = UV^{\dagger}$.
This is shown in Fig.~\ref{fig: update}(d), and is achieved with a computational cost $\mathcal{O}(d^{6})$.

\subsection{Contracting the environment}
\label{sec: environment}

\begin{figure}[t]
    \centering
    \includegraphics[width=\linewidth]{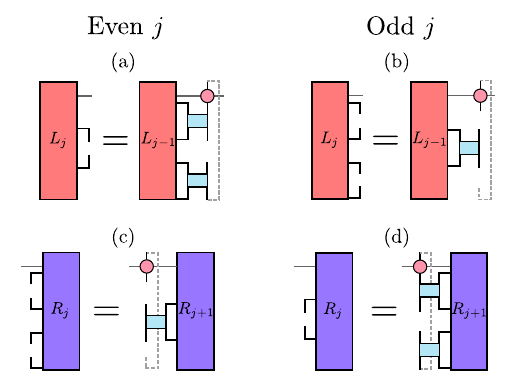}
    \caption{\textbf{Contracting the environment.}
    Building the environment blocks, illustrated for a circuit depth $M = 3$ circuit.
    (a, b) Building the left block $L_{j}$ from the previous $L_{j-1}$ for even and odd $j$ respectively.
    (c, d) Building the right block $R_{j}$ from the previous $R_{j+1}$ for even and odd $j$ respectively.
    }
    \label{fig: environment}
\end{figure}

As is often done in TN methods, it is convenient and efficient to recycle partially contracted networks \cite{Schollwock2011}.
For this method, it will be useful to save each of the blocks $L_{j}$ and $R_{j}$ to memory for two reasons.
The first is the blocks $L_{k-1}$ and $R_{k+1}$ will be the same for each of the $M$ gates which act on site $k$.
They can be calculated and then recycled to update each of the $M$ gates.
The second reason is that the blocks $L_{j}$ and $R_{j}$ can be used iteratively to calculate the blocks at all sites $j$.
That is, $L_{j}$ can be calculated using $L_{j-1}$ and $R_{j}$ can be calculated using $R_{j+1}$.
This procedure is illustrated in Fig.~\ref{fig: environment} for a circuit depth $M = 3$, although the exact calculations will differ for other circuit depths.
The left column shows how to construct the blocks for even $j$, and the right column for odd $j$.
The block $L_{j-1}$ is contracted with all the gates which act on both sites $j-1$ and $j$, along with the adjoint of the MPO tensor at site $j$ to form the block $L_{j}$. 
Similarly, the block $R_{j+1}$ is contracted with the gates which act on both sites $j$ and $j+1$, along with the adjoint of the MPO tensor at site $j$ to give $R_{j}$.
The cost of expanding such blocks is at most $\mathcal{O} \left(\chi^{2} d^{2 + 2\lceil M/2 \rceil} + \chi M d^{4 + 2\lceil M/2 \rceil} \right)$, and storing to memory is at most $\mathcal{O} \left( \chi d^{2\lceil M / 2 \rceil} \right)$ per block.
See Table.~\ref{table: costs} for a complete description of all the computational costs in our method.

\begin{table}[t]
    \def\arraystretch{1.25}% 
    \begin{centering}
        \begin{tabular}{|@{}c@{}|}
            \hline
            \rowcolor{lightgray!40}
            \textbf{Estimating $\bm{U(\Delta t)}$ as an MPO}
            \\
            \hline
            $\mathcal{O}(N \Delta t \, \Delta \tau^{-1} \chi^{3} d^{8})$
            \\ 
            \rowcolor{lightgray!40}
            \hline
            \textbf{Updating the environment}
            \\
            \hline
            $\mathcal{O}(\chi^{2}d^{2 + 2\lceil M/2 \rceil}) + 
            \mathcal{O}(\chi M d^{4 + 2\lceil M/2 \rceil})$
            \\ 
            \hline \rowcolor{lightgray!40}
            \textbf{Updating the gates (per gate)}
            \\
            \hline
            $\mathcal{O}(\chi^{2}d^{2 + 2\lceil M/2 \rceil}) + 
            \mathcal{O}(\chi M d^{4 + 2\lceil M/2 \rceil})$
            \\ 
            \hline \rowcolor{lightgray!40}
            \textbf{Cost per sweep}
            \\
            \hline
            \qquad\qquad
             $\mathcal{O}(N\chi^{2}d^{2 + 2\lceil M/2 \rceil}) + \mathcal{O}(N\chi M d^{4 + 2\lceil M/2 \rceil})$ \qquad\qquad
            \\
            \hline
        \end{tabular}
    \end{centering}
    \caption{\label{table: costs} The computational complexity for each of the steps in the optimisation method.
    $N$ is the number of lattice sites; $d$ is the local dimension of the system; $\Delta t$ is the time step of unitary $U(\Delta T) = e^{-i\Delta t\hat{H}}$; $\Delta \tau$ is the time step in Trotterization; $\chi$ is the bond dimension of the MPO $U(\Delta t)$; $M$ is the depth of the BW circuit.}
\end{table}

\section{Results} \label{sec: results}
We now benchmark our approach for numerous models of interest for system sizes $N = 8$ to $512$.
To assess the quality of the dynamics, we first measure the error of the BW circuit with respect to the high-accuracy MPO approximation, and investigate violations of conserved quantities.
We then calculate the number of CNOT layers needed to implement the circuits on a quantum computer, which we compare to Trotterization.
Finally, we investigate the convergence properties of our method to demonstrate its practicality.

\begin{figure}[t]
    \centering
    \includegraphics[width=\linewidth]{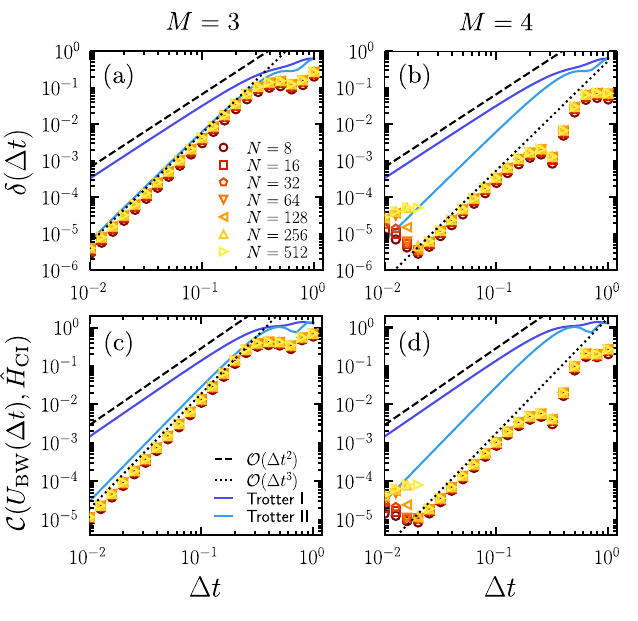}
    \caption{\textbf{Error scaling for the CI model.}
    (a, b) The error density of the BW circuit with respect to the MPO approximation, $\delta(\Delta t)$ and (c, d) the Frobenius norm of the commutator with the Hamiltonian, \er{commutator_conserved}.
    The data points show the results from variational optimisation for system sizes $N = 8$ to $512$.
    The solid lines show the same results for first and second order Trotterization with $N = 8$.
    The dashed lines show the scaling $\mathcal{O}(\Delta t^{2})$ and the dotted lines show the scaling $\mathcal{O}(\Delta t^{3})$.
    The left column is for depth $M = 3$, and the right column is for depth $M = 4$.
    All results are for $g = -0.75$.
    }
    \label{fig: scaling_ci}
\end{figure}

\subsection{Error metrics}

We first consider the performance of the unitary circuits with respect to the time step, $\Delta t$.
One instructive measure is the error density of the circuit with respect to the MPO
\footnote{In practice, we expect that $\Tr [ U^{\dagger}_{\rm MPO}(\Delta t) U_{\rm BW} (\Delta t)] = (2 - \delta(\Delta t)^{2})^N$; \er{error_density} follows directly from this.},
\beq
    \delta(\Delta t) = \sqrt{2 - {\rm Re}\left(\Tr \left[ U^{\dagger}_{\rm MPO}(\Delta t) U_{\rm BW} (\Delta t) \right]\right)^{1/N}}.
    \label{error_density}
\eeq
For small $\Delta t$, the error density for first and second order Trotterization will scale as $\mathcal{O}(\Delta t^{2})$ and $\mathcal{O}(\Delta t^{3})$ respectively; this will provide both a qualitative and quantitative criterion to compare our results to.
It is also important to note that we must achieve a scaling better than $\mathcal{O}(\Delta t)$ to have meaningful results:
if the error of the gates goes as $\mathcal{O}(\Delta t)$, then the total error for time $t$ is $\mathcal{O}(\Delta t)$, and using a smaller $\Delta t$ will yield no improvements.

A second quantity to consider is the Frobenius norm of the commutator of the BW circuit with some conserved quantities, $\hat{O}$,
\beq
    \mathcal{C}(U_{\rm BW}(\Delta t), \hat{O}) = \frac{|| [U_{\rm BW}(\Delta t), \hat{O}] ||_{\rm F}}{|| \hat{O} ||_{\rm F}},
    \label{commutator_conserved}
\eeq
which is normalised by the Frobenius norm of the operator $\hat{O}$.
Note that unlike $\delta(\Delta t)$ --- which calculates the error with respect to a high accuracy {\em approximation} of the true time propagator --- this quantity can be calculated {\em exactly} if $\hat{O}$ has an efficient MPO representation.
For example, \er{commutator_conserved} can be used with any local Hamiltonian, $\hat{O} = \hat{H}$, to investigate the conservation of energy.

\subsubsection{Cluster Ising model}

The first model we consider is a cluster Ising (CI) model, 
\begin{multline}
    \hat{H}_{\rm CI} = -(1+g)^{2} \sum_{j=1}^{N} \hat{X}_{j} - 2(1-g^{2})\sum_{j=1}^{N-1} \hat{Z}_{j}\hat{Z}_{j+1} 
    \\
    + (g-1)^{2}\sum_{j=1}^{N-2} \hat{Z}_{j}\hat{X}_{j+1}\hat{Z}_{j+2},
    \label{H_ci}
\end{multline}
where $g \in [-1, 1]$ is a tuning parameter which interpolates between a symmetry protected topological phase for $g < 0$ and topologically trivial phase for $g > 0$ \cite{Verresen2017, Verresen2018}. 
The operators $\hat{X}_{j}$ and $\hat{Z}_{j}$ are the usual Pauli operators for two-level systems, acting at the lattice site $j$. 
This choice of Hamiltonian is motivated by recent works \cite{Smith2022} which show how the ground state of \er{H_ci} can be prepared on quantum simulators.

We apply our variational approach to \er{H_ci}.
We choose a random initial unitary circuit with each gate random, but close to identity.
For $g \leq 0$, we find the initial circuit plays little role in convergence: the method is able to consistently converge to some optimal value.

Figure~\ref{fig: scaling_ci} show the results for the symmetry protected phase with depths $M = 3$ (left panels) and $M = 4$ (right panels).
The top panels show the error density, $\delta(\Delta t)$, and the bottom panels show the error in the conservation of energy, $\mathcal{C}(U_{\rm BW}(\Delta t), \hat{H}_{\rm CI})$.
Similar results for the trivial phase are found in App.~\ref{app: errors_ci}.
In each case, we find the minimal depth needed to obtain at least a scaling $\mathcal{O}(\Delta t^2)$ is $M \geq 3$.
Interestingly, there are no circuit depths which yield results comparable to first order Trotterization, shown by the darker solid line (which has scaling $\mathcal{O}(\Delta t^{2})$, shown by the dashed line).
Instead, the circuit depth $M = 3$ has scaling $\mathcal{O}(\Delta t^3)$ (dotted line), matching the results of second order Trotterization (lighter solid line).
Indeed, we find the variational optimisation is even able to give prefactor improvements to Trotterization, but with a drastic reduction in the number of CNOT layers required, see Sec.~\ref{sec: cnot}.

We also observe $\mathcal{O}(\Delta t^3)$ behaviour for the circuit depth $M = 4$. 
However, it should be noted that the extra BW layer yields an order of magnitude improvement in the error and conservation of energy.
Furthermore, we observe a dip in the error around $\Delta t \sim 0.4$ which is consistent with system size.
While we are not able to offer an explanation for this behaviour, the key observation is that additional layers might provide significant improvement.
Notice that, for small times $\Delta t \sim 10^{-2}$, the optimisation method can struggle to converge to the expected scaling.
Nevertheless, the errors are on the order of numerical precision.

\subsubsection{The PXP model}
\begin{figure}[t]
    \centering
    \includegraphics[width=\linewidth]{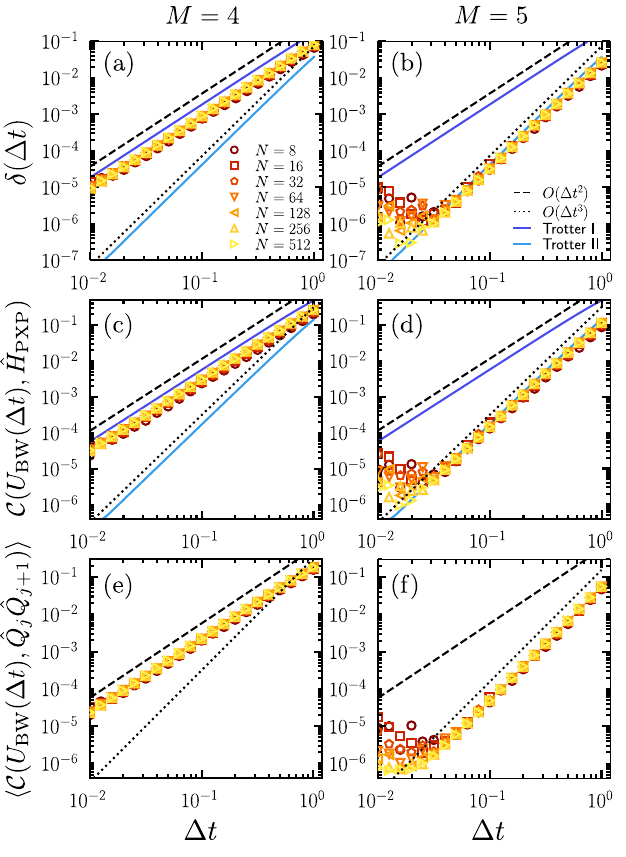}
    \caption{\textbf{Error scaling for the PXP model.}
    Error metrics for the PXP model with circuit depths $M = 4$ (left panels) and $M = 5$ (right panels).
    (a, b) The error density $\delta(\Delta t)$ as a function of $\Delta t$.
    (c, d) Violation of the conservation of energy, $\mathcal{C}(U_{\rm BW}(\Delta t), \hat{H}_{\rm PXP})$.
    (e, f) Violation of the conserved quantities, $\braket{\mathcal{C}(U_{\rm BW}(\Delta t), \hat{Q}_{j}\hat{Q}_{j+1})}$.
    %The data points are the results of variational optimization, with a random initial seed and warm-up strategy where the time is annealed from large-to-small, see the main text for details.     
    %The solid lines show the same quantity for first and second order Trotterization for a system size of $N = 8$.
    %The dashed and dotted lines show $\mathcal{O}(t^{2})$ and $\mathcal{O}(t^{3})$ respectively.
    The data points and curves are as defined in Fig.~\ref{fig: scaling_ci}.}
    \label{fig: scaling_pxp}
\end{figure}

The second model we consider is the PXP model \cite{Fendley2004, Lesanovsky2011, Turner2018},
\beq
    \hat{H}_{\rm PXP} = \hat{X}_{1}\hat{P}_{2} + \sum_{j=2}^{N-1} \hat{P}_{j-1}\hat{X}_{j}\hat{P}_{j+1} + \hat{P}_{N-1}\hat{X}_{N},
    \label{Hpxp}
\eeq
where $\hat{P}_{j} = \frac{1}{2}(\hat{\mathds{1}} - \hat{Z}_{j})$ is the local projection onto the spin down state, and $\hat{\mathds{1}}$ is the identity operator.
The PXP model is an idealised {\em kinetically constrained model} for realising the dynamics of cold atom platforms \cite{Browaeys2020} in a strong coupling limit, and has recently received much attention due to its connection to quantum many-body scars \cite{Turner2018, Turner2018b}.
Furthermore, unlike the CI model considered in the previous section, the PXP model is non-integrable \cite{Turner2018}.

Figure~\ref{fig: scaling_pxp} shows the results of our optimisation method.
We use the variational approach with an annealing strategy which reduces the time $\Delta t$ of the MPO from an initial time of $\Delta t = 1.0$ to the target time.
The first column is for depth $M = 4$ (which was the minimal depth needed to reach order $k = 1$), and the second column shows depth $M = 5$.
The first row shows the error density $\delta(\Delta t)$.
For a depth of $M = 4$, we are able to obtain an $\mathcal{O}(\Delta t^{2})$ scaling, with prefactor improvements on first-order Trotterization.
For a depth of $M = 5$, this can be improved to an $\mathcal{O}(\Delta t^{3})$ scaling, matching the results of second-order Trotterization.
Notice that, as was the case for the CI model, the variational approach can sometimes struggle to converge for small $\Delta t$.

The second row of Fig.~\ref{fig: scaling_pxp} shows the error in the conservation of energy, $\mathcal{C}(U_{\rm BW}(\Delta t), \hat{H}_{\rm PXP})$.
This value scales in the same way as the error density, and also closely matches the results of Trotterization.
The PXP model also conserves the quantities $\hat{Q}_{j}\hat{Q}_{j+1}$ for all $j$, with $\hat{Q}_{j} = \hat{\mathds{1}} - \hat{P}_{j}$.
We calculate how well this quantity is conserved through the lattice site average 
\begin{multline}
    \braket{\mathcal{C}(U_{\rm BW}(\Delta t), \hat{Q}_{j}\hat{Q}_{j+1})} =
    \\ (N-1)^{-1} \sum_{j=1}^{N-1} \mathcal{C}(U_{\rm BW}(\Delta t), \hat{Q}_{j}\hat{Q}_{j+1}),
\end{multline}
shown in the bottom row of Fig.~\ref{fig: scaling_pxp}.
The error in these conserved quantities is qualitatively the same as the error density and the error in the conservation of energy.
However, it is important to note that the conserved quantities of the PXP models are a result of the kinetic constraints in \er{Hpxp}, and are {\em exactly} conserved by Trotterization.

\subsubsection{The next-nearest-neighbour Ising model}
\begin{figure}[t]
    \centering
    \includegraphics[width=\linewidth]{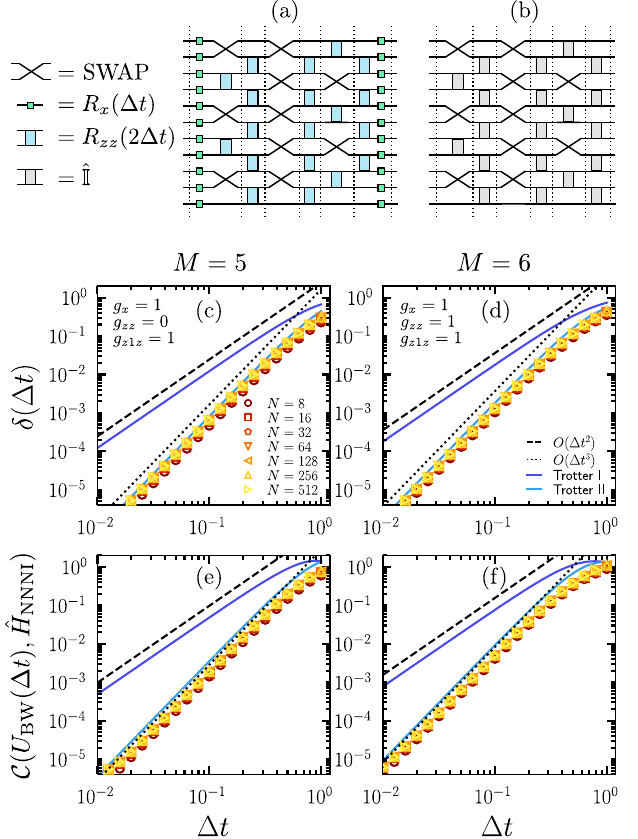}
    \caption{\textbf{Error scaling for the NNNI model.}
    (a) A second order trotter decomposition for the NNNI model can be written as a BW circuit composed of SWAP gates, $R_{x}(\Delta t) = e^{-i\Delta t\hat{X}/2}$ gates and $R_{zz}(\Delta t) = e^{-i\Delta t\hat{Z}\otimes \hat{Z}}$ gates. The dashed lines show where gates can be compressed to give one layer of the BW circuit.
    (b) As an initial guess for variational optimisation with circuit depth $M = 6$, we use a similar circuit where the SWAP gates are placed in the same positions as (a). The remainder of the gates are set to identity.
    For depth $M = 5$, the initial guess is the same, but with the final layer removed.
    (c, d) The error density, $\delta(\Delta t)$.
    (e, f) The error in the conservation of energy, $\mathcal{C}(U_{\rm BW}(\Delta t), \hat{H}_{\rm NNNI})$.
    Panels (c) and (e) show results for $g_{x} = g_{z1z} = 1$ and $g_{zz} = 0$ with circuit depth $M = 5$, while panels (d) and (f) show results for $g_{x} = g_{zz} = g_{z1z} = 1$ and $M = 6$.
    The data points and curves are as defined in Fig.~\ref{fig: scaling_ci}.
    }
    \label{fig: scaling_nnni}
\end{figure}

The final model we consider is the next-nearest-neighbour Ising (NNNI) model,
\begin{multline}
    \hat{H}_{\rm NNNI} = g_{x} \sum_{j=1}^{N} \hat{X}_{j} + g_{zz} \sum_{j=1}^{N-1} \hat{Z}_{j}\hat{Z}_{j+1} 
    \\
    + g_{z1z} \sum_{j=1}^{N-2} \hat{Z}_{j}\hat{Z}_{j+2}.
    \label{H_nnni}
\end{multline}
Its second order Trotterization can be written as a BW unitary circuit through a decomposition into SWAP gates and $R_{zz}(\Delta t) = e^{-i\Delta t\hat{Z}\otimes\hat{Z}}$ gates, as illustrated in Fig.~\ref{fig: scaling_nnni}(a).
By grouping gates together (shown by the dashed lines), the second order decomposition becomes a BW circuit with depth $M = 6$.
For $g_{zz} = 0$, this can be done for $M = 5$ by removing the gates responsible for the nearest-neighbour interactions in \er{H_nnni}.

Unlike the two previous models, we find that the optimisation for the NNNI model is highly susceptible to becoming stuck in local minima.
Because of this, optimising the circuits to achieve a scaling that beats $\mathcal{O}(\Delta t)$ is extremely difficult, and only happens on rare occasions for some choice of initial guess. 
This effect becomes more apparent for larger system sizes.
As such, this model serves as an interesting example where the solution is known explicitly, but the variational optimisation method is unable to reliably obtain a good result.
Instead, we propose an initial guess which uses the structure of the circuit in Fig.~\ref{fig: scaling_nnni}(a), but only for the SWAP gates, see Fig.~\ref{fig: scaling_nnni}(b).
All other gates are set to identity.
With this choice of initial guess, we find the error metrics converge to those of the second order Trotterization solution, see Figs.~\ref{fig: scaling_nnni}(c, d) for the error density, and Figs.~\ref{fig: scaling_nnni}(e, f) for the conservation of energy.
These results could motivate the future investigation of strategies to systemically suggest more optimal choices of initial guesses.
Notice that in this instance, we observe little improvement when compared to the results of Trotterization.

\subsection{Number of CNOT layers}
\label{sec: cnot}

While two-qubit gates are required to create entanglement, they are often the operations which take the longest to implement on quantum computers, and thus are most responsible for decoherence.
For many popular platforms, such as IBM's quantum processors, they are implemented using CNOT gates.
Because of this, it is essential to implement quantum dynamics using as few CNOT layers as possible.
We discuss how both the BW circuits and Trotterized circuits can be implemented using just single qubit operations and CNOT gates in App.~\ref{app: decompositions}, but summarise the results here and in Tab.~\ref{table: cnots}.

\begin{table}[t]
    \bgroup
    \def\arraystretch{1.25}% 
    \begin{centering}
        \begin{tabular}{c | c | c | c }
            \hline\hline
            \multicolumn{2}{c |}{\textbf{Circuit}} & \textbf{CNOT Layers} & \textbf{Error,} $\delta(\Delta t=0.1)$ \\
            \hline
            \multirow{3}{*}{CI} & Trotter $k=1$ & 16 & $3.03 \times 10^{-2}$ \\
            & Trotter $k=2$ & 28 & $6.32 \times 10^{-3}$  \\
            & Brickwall $k=2$ & 9 & $3.03 \times 10^{-3}$ \\
            \hline
            \multirow{4}{*}{PXP} & Trotter $k=1$ & 42 & $1.87 \times 10^{-3}$ \\
            & Trotter $k=2$ & 70 & $4.23 \times 10^{-5}$  \\
            & Brickwall $k=1$ & 12 & $6.67 \times 10^{-4}$ \\
            & Brickwall $k=2$ & 15 & $3.06 \times 10^{-5}$ \\
            \hline
            \multirow{3}{*}{NNNI} & Trotter $k=1$ & 13 & $1.78 \times 10^{-2}$ \\
            & Trotter $k=2$ & 15 & $1.81 \times 10^{-3}$  \\
            & Brickwall $k=2$ & 18 & $1.04 \times 10^{-3}$ \\
            \hline\hline
        \end{tabular}
    \end{centering}\egroup
    \caption{\label{table: cnots} 
    We compare the circuits from Trotterization to ones from our optimisation method.
    The maximal number of CNOT layers required to implement a single time step of the Trotterized circuits and BW circuits for any system size is shown in the second column.
    The final column shows the error density $\delta(\Delta t)$ (rounded to three significant figures), for time step $\Delta t = 0.1$ and system size $N = 8$.
    Results are shown for Cluster Ising (CI) model, the PXP model and a next-nearest-neighbour Ising (NNNI) model.
    The order of the approximation, $k$, is given for each circuit. Note that the error density scales as $\mathcal{O}(\Delta t^{k+1})$ for small $\Delta t$.
    The CI model is for $g = -0.75$ and the NNNI model is for $g_{x} = g_{zz} = g_{z1z} = 1$.}
\end{table}

Each layer of BW unitary gates can be decomposed into a quantum circuit with three layers of CNOT gates.
It follows that a BW circuit of depth $M$ has at most $3M$ layers of CNOT gates.
The Trotterized circuits for each of the models requires more care.
The first and second order Trotterized circuits for the CI model can be implemented with 16 CNOT layers and 28 CNOT layers respectively.
On the contrary, a BW circuit of depth $M = 3$ gives a second order scaling while requiring only 9 CNOT layers.
This yields an enhancement factor of $\sim 3.1$ when compared to second order Trotterization, and can even be implemented with fewer CNOT layers than first-order Trotterization.
A circuit depth of $M = 4$ also has a second order scaling, but with order-of-magnitude improvements on second order Trotterization.
This can be implemented with $12$ CNOT layers.

For the PXP model, we found that circuit depths $M = 4$ and $M = 5$ gave first and second order scaling respectively, requiring a total of $12$ and $15$ layers of CNOTs.
On the contrary, first order Trotterization requires 42 CNOT layers (assuming there are only CNOT connections between neighbouring qubits), giving an enhancement factor of $3.5$.
Similarly, second order Trotterization requires 70 CNOT layers, giving an enhancement factor of $\sim 4.6$.

The final model we considered was the NNNI model.
Here, both the BW circuits and the Trotterized circuit gave a second order scaling with the same number of layers of two-qubit gates.
Naively, one might assume they can be implemented with the same number of CNOT gates.
However, it is important to note that the two-qubit gates in the Trotterized circuit only include SWAP gates and $R_{zz}(\Delta t)$ gates, while the classically optimised BW circuits allow for arbitrary two-qubit gates.
Any general two-qubit gate (including SWAP gates) can be implemented with three CNOT gates, while the $R_{zz}(\Delta t)$ requires only two CNOTs.
Because of this, the Trotterized circuit can be implemented with fewer CNOT layers than the classically optimised circuits.
For $g_{zz} = 0$ ($g_{zz} \neq 0$), the optimised BW circuit requires $15$ ($18$) CNOT layers, while the Trotterized circuit requires only $13$ ($15$) layers.
Note that, while the BW circuit requires more CNOT layers to implement the dynamics, it yields an improvement in the error density, see Tab.~\ref{table: cnots}.

\subsection{Convergence of optimisation}
\begin{figure}[t]
    \centering
    \includegraphics[width=\linewidth]{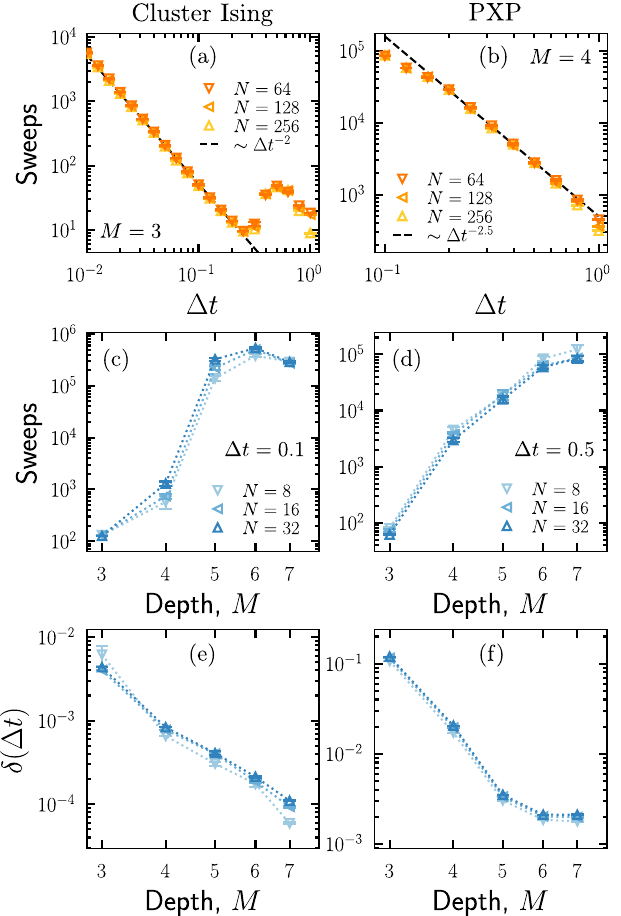}
    \caption{\textbf{Convergence of variational optimisation.}
    The left panels are for the CI model with $g = -0.5$, and the right panels are for the PXP model.
    (a, b) The number of sweeps required to converge for a fixed circuit depth, $M$, as a function of time, $\Delta t$.
    Data is shown for $N = 64, 128, 256$ with a circuit depth of $M = 3$ for the CI model, and $M = 4$ for the PXP model.
    For small times, the number of sweeps required to converge goes as some power law, $\Delta t^{-\alpha}$, with $\alpha = 2$ for the CI model and $\alpha \approx 2.5$ for the PXP model, shown by the dashed lines. 
    (c, d) The number of sweeps required to converge for a fixed time, $\Delta t$, as a function of circuit depth, $M$.
    Data is shown for $N = 8, 16, 32$ with a time of $\Delta t = 0.1$ for the CI model, and $t = 0.5$ for the PXP model.
    (e, f) The error density, $\delta(\Delta t)$, averaged over all runs for the same systems as panels (c) and (d).
    Each data point is the mean of $100$ independent runs.
    }
    \label{fig: benchmarking}
\end{figure}

We now benchmark the performance of the variational optimisation.
In particular, we test the number of sweeps required for convergence.
After each sweep, $l$, we measure the Frobenius norm of the difference between our BW circuit ansatz and the MPO approximation, $\mathcal{F}_{l}^{(1)} = \sqrt{\mathcal{F}_{l}^{(2)}}$, where $\mathcal{F}_{l}^{(2)}$ is \er{cost} after the $l$-th sweep.
We define convergence to be when the relative difference in the Frobenius norm between successive sweeps falls below some value $\epsilon$ 
\beq
    \frac{\mathcal{F}_{l-1}^{(1)}- \mathcal{F}_{l}^{(1)}}{\mathcal{F}_{l-1}^{(1)}} < \epsilon.
\eeq
In practice, we use $\epsilon = 10^{-6}$.

We first check the number of sweeps required for convergence as a function of time, with a fixed depth $M$, which we choose to be the minimum depth that is required to yield an error density that scales better than $\mathcal{O}(\Delta t)$.
For the CI model, this is a circuit depth of $M = 3$, and for the PXP model, a circuit depth of $M = 4$.
This is shown in Figs.~\ref{fig: benchmarking}(a, b) for the CI model and PXP model respectively with system sizes $N = 64, 128, 256$.
Each data point shows the average over one hundred independent runs.
Notice that in both cases, the number of sweeps required to converge looks to be independent of system size.
Furthermore, for small times $\Delta t \lesssim 1$, the number of sweeps needed to converge appears to go as a power law, $O(\Delta t^{-\alpha})$; for the CI model, there is an exponent $\alpha = 2$, and for the PXP model, an exponent of $\alpha \approx 2.5$.
While we cannot offer a compelling explanation for the exact values of these exponents, we speculate they are related to the transport properties of the models. 
Nevertheless, the key point is that this scales sub-exponentially. 
Notice that the number of sweeps deviate from the scaling at large times, where the error density no longer scales as a polynomial (c.f. Fig.~\ref{fig: scaling_ci}).
Furthermore, while it might seem counter-intuitive that the number of 
sweeps needed for convergence decreases with larger time step $\Delta t$, note that the accumulated error for evolving to a total time $t$ with a small time step $\Delta t$ goes as $\mathcal{O}(t\Delta t^{k})$, with $k = 2$ for the CI model and $k = 1$ for the PXP model.
It follows that decreasing the time step will result in a more accurate evolution.

We also investigate convergence for a fixed time step, $t$, and variable circuit depth, $M$.
Figures~\ref{fig: benchmarking}(c, d) show the average number of sweeps required for convergence as a function of $M$ for system sizes $N = 8, 16, 32$ for the CI model and PXP model respectively.
While there is no obvious trend, it is clear that the number of sweeps needed for convergence scales subexponetially in the circuit depth.
However, we emphasise that while the number of sweeps might not scale exponentially, the cost required for updating unitary gates does.
We also show the average error density for the converged circuits in Figs.~\ref{fig: benchmarking}(e, f), for the same systems as before.
Notice that while increasing the circuit depth consistently provides an improvement to the error, it looks to be diminishing.
The most substantial improvements are found when increasing the circuit depth changes the scaling of the error density for small times, i.e., increasing the depth $M$ increases the exponent $k$ for $\delta(\Delta t) \sim \mathcal{O}(\Delta t^{k+1})$.
This is evident from panel (f), where circuit depths $M = 3, 4, 5$ gives orders $k = 0, 1, 2$ respectively.
In practice, we find that is it more effective to optimise for a shallow circuit which also yields a desirable scaling.

\section{Optimal time step} \label{sec: ibm}
We have explained how to optimise a BW circuit to efficiently simulate quantum dynamics for some time $\Delta t$, which can then be used repeatedly to achieve some target time $t$ that is unobtainable on a classical computer.
However, it is not immediately clear what time step $\Delta t$ should be chosen to simulate the dynamics.
On one hand, we have shown that choosing a smaller time step can reduce the {\em algorithmic error} for evolving to time $t$.
Indeed, if a BW circuit of depth $M$ gives a $k$th order approximation (e.g. $M = 3$ for the CI model has order $k = 2$), then the cost of evolving to time $t$ is $\mathcal{O}(t\Delta t^{k})$.
This error is illustrated by the dashed red line in Fig.~\ref{fig: optimal_sketch}.
This entices us to pick a small time step, resulting in a deeper quantum circuit.
On the other hand, quantum computers are never completely isolated from their environment.
This introduces errors which are native to the {\em device}.
We assume that this error is exponential in the circuit depth, as shown by the blue dashed line in Fig.~\ref{fig: optimal_sketch}.
This indicates that our simulations would benefit from a shallower quantum circuit.
The total simulation error is a combination of the two, shown by the black curve, and the optimal time step is the one which minimises this error.
In what follows, we explain how to estimate each of these errors to guide us in determining an optimal time step.
We note that a similar analysis was done for Trotterization in Ref.~\cite{Kuper2022}

\subsection{Measures of fidelity on quantum devices}

One way to estimate the error native to a quantum computer is to simulate a unitary BW circuit, followed by the same circuit, but in reverse.
For a perfect quantum computer, the resulting operation is the identity matrix, $UU^{\dagger} = \hat{\mathds{1}}$, and the initial state will be retrieved with certainty.
However, in practice, each layer of the quantum circuit will incur some random error native to the device, and the circuit described above will not yield the identity matrix.
This allows us to estimate a layer-dependent error of the device through the {\em fidelity} of the state.

\begin{figure}[t]
    \centering
    \includegraphics[width=0.9\linewidth]{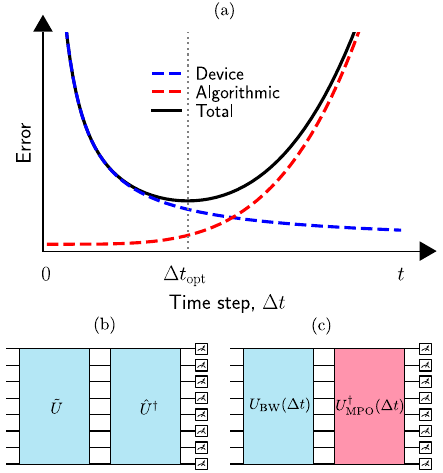}
    \caption{\textbf{Optimal time step.}
    (a) The optimal time step for quantum simulation can be determined through considerations of the errors native to the quantum device (blue dashed line) and the algorithmic errors from the optimisation of BW circuits (red dashed line).
    Small time steps, $\Delta t$, require a large circuit depth to reach time $t$, and accumulate a large error from the device.
    On the contrary, the BW circuit is unable to well approximate the true time evolution operator for large time steps.
    The optimal time step, $\Delta t_{\rm opt}$, is determined through a combination of the two.
    (b) The native error of a quantum computer can be measured by running a BW circuit $U$ forwards and then in reverse, and taking measurements.
    In practice, the quantum computer will add noise to the evolution, and we can consider this the same as evolving by some noisy circuits $\tilde{U}$ and $\hat{U}^{\dagger}$.
    The error can be calculated from the discrepancy between the initial state and the measurements; see the main text for details.
    (c) The algorithmic error can be calculated with a similar set-up, where this time we evolve using the BW circuit, $U_{\rm BW}(\Delta t)$, but then evolve using the adjoint of the MPO, $U_{\rm MPO}^{\dagger}(\Delta t)$.}
    \label{fig: optimal_sketch}
\end{figure}

Let us assume the system is prepared in a state such that each qubit is an eigenstate of the $Z$-basis, e.g., $\ket{\psi} = \bigotimes_{j=1}^{N} \ket{\sigma_{j}}$ with $\sigma_{j} = \pm 1$.
We can then evolve the system using a unitary BW circuit, followed by the same circuit in reverse.
Following this, one can then take a measurement for each qubit in the $Z$-basis.
This is shown in Fig.~\ref{fig: optimal_sketch}(b).
Note that, in practice, the error might depend on the unitary circuit, and one should use a circuit which resembles the circuits that simulate the dynamics of the Hamiltonian.
The fidelity of the device for the total circuit goes as
\beq
    F^{\rm d} = \frac{1}{2^N}\bigg<\sum_{\bm \sigma} \left|\braket{{\bm \sigma} | \hat{U}^{\dagger} \tilde{U} | {\bm \sigma}}\right|^{2} \bigg>_{\rm noise},
    \label{fidelity}
\eeq
where $\ket{\bm \sigma} = \ket{\sigma_{1} \cdots \sigma_{N}}$, and $\tilde{U}$ and $\hat{U}$ are the unitary circuit $U$ subject to two independent realisations of the random noise of the quantum computer.
The $\left< \cdot \right>_{\rm noise}$ indicates an average over random noise.
In practice, one can estimate \er{fidelity} by simulating the circuit shown in Fig.~\ref{fig: optimal_sketch}(b) a finite number of times (sometimes referred to as {\em shots}) each with a random initial state.

While the fidelity provides an unbiased way to estimate the native errors of quantum simulation, in practice it is exponentially difficult to measure in the number of qubits: the number of shots needed for a measurement with some desired accuracy goes as $\mathcal{O}(e^{N})$ \cite{Tilly2022}.
Typically, one is interested in studying the evolution of local quantities, such as the expectation value of local observables.
As such, it could be beneficial to instead determine an error using only local measurements.
As before, we can prepare the initial state of the system to be a product state of $Z$-basis eigenstates, and then evolve it forwards and backwards. 
We then take measurements in the $Z$-basis.
However, this time, we calculate a {\em local fidelity},
\beq
    \tilde{F}^{\rm d} = \frac{1}{N2^N}
    \bigg<\sum_{\bm \sigma} \sum_{j=1}^{N} \braket{{\bm \sigma} | \hat{U}^{\dagger} \tilde{U}\hat{\sigma}_{j}
    \tilde{U}^{\dagger} \hat{U} | {\bm \sigma}} \bigg>_{\rm noise},
    \label{fidelity_opt}
\eeq
where $\hat{\sigma}_{j} = (\hat{\mathds{1}} + {\rm sgn}(\sigma_{j}) \hat{Z}^{j}) / 2$ is a projector onto the state $\sigma_{j}$ (i.e. the $j$th spin in the initial state ${\bm \sigma}$).
Notice that \er{fidelity_opt} can be measured with the same set up as \er{fidelity}.
Furthermore, the definition can easily be extended to consider multiple-site observables.

\begin{figure}[t]
    \centering
    \includegraphics[width=\linewidth]{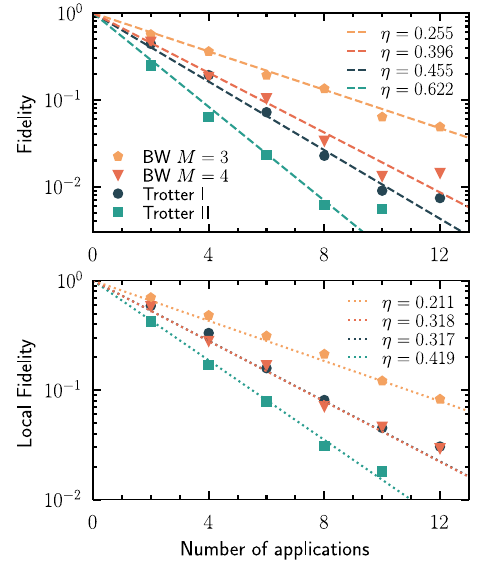}
    \caption{\textbf{Fidelity of the quantum device.}
    We run the forwards-backwards set-up from Fig.~\ref{fig: optimal_sketch}(b) to measure the errors on the device {\texttt ibm\_brisbane}, which was accessed on the 11th of November 2023.
    Results are for the CI model with $g = -0.5$, $N = 5$ qubits (each starting from spin down), and time step $\Delta t = 0.25$.
    The top panel shows the normalised fidelity as a function of the total number applications of each circuit.
    The bottom panel shows the same, but for the normalised local fidelity.
    Data is shown for first order Trotterization, second order Trotterization, and optimised BW circuits with depths $M = 3, 4$.
    The dashed/dotted line shows the exponential function with exponent $-\eta$, with $\eta$ given in the legends.
    Each data point is averaged over $8192$ shots.
    }
    \label{fig: ibm_brisbane}
\end{figure}

We demonstrate our approach in Fig.~\ref{fig: ibm_brisbane} using the {\texttt ibm\_brisbane} device, where we simulate the dynamics of the CI Ising for $N = 5$ qubits using a time-step of $\Delta t = 0.25$, with first and second order Trotterization, and with BW circuits with depths $M = 3, 4$.
We first evolve forwards in time by multiple applications of time step $\Delta t$, and then backwards in time, with the same number of applications of time step $-\Delta t$, as described in the set-up above.
While technical limitations only allowed us to run the circuits with the initial state where all spins are $\sigma_{j} = -1$, in practice, one should average over many initial conditions.

The top panel shows the normalised fidelity $(F^{d} - 2^{-N}) / (1 - 2^{-N})$ as a function of the number of total applications of each circuit (which is twice the number of forward applications).
The bottom panel shows the same, but for the normalised local fidelity, $2\tilde{F}^{d} - 1$ 
\footnote{The fidelities are normalised such that the value of the fidelity is one initially, and is zero after all memory of the initial conditions is lost (and the state of the system can be assumed to be random).}.
The dashed/dotted lines show the exponential fit $e^{-\eta x}$ for each type of circuit, where $x$ is the number of applications.
It is clear that the fidelity decays approximately exponentially for both measures.
While the number of applications of each circuit is the same, it is important to note that each circuit will differ in its quantum circuit depth.
We expect the most significant source of error to come from the number of CNOT layers.
Indeed, the BW circuit with $M=3$ outperforms both first and second order Trotterization while attaining an error density on par with second order Trotterization (c.f. Fig.~\ref{fig: scaling_ci}). 
The BW circuit with depth $M=4$ has a native error similar to that of first order Trotterization, but with an error density which outperforms second order Trotterization by a significant prefactor.

\subsection{Measures of fidelity for classically optimised circuits}

The error density from Sec.~\ref{sec: results} allows us to understand the algorithmic error of a single time step with respect to the true evolution operator.
However, it cannot be used comparatively with the fidelity measures of the previous section.
We can instead use a forwards and backwards scheme as was done in the previous section, but this time, with the forwards evolution done with the BW circuit approximation, and the backwards evolution done with the true dynamics, $U^{\dagger}(\Delta t)$, which in practice we approximate with an MPO, $U(\Delta t) \approx U_{\rm MPO}(\Delta t)$.
That is, we calculate 
\beq
    F_{\Delta t}^{\rm a} = \frac{1}{2^{N}}
    \sum_{\bm \sigma} |\braket{{\bm \sigma} | U_{\rm MPO}^{\dagger}(\Delta t) U_{\rm BW}(\Delta t) | {\bm \sigma}}|^{2}.
    \label{fidelity_algorithmic}
\eeq
This is illustrated in Fig.~\ref{fig: optimal_sketch}(c).

While we are unable to bound the fidelity using the error density (or the Frobenius norm), it is possible to bound the fidelity using the operator norm \cite{Yi2022}.
If we have $|| U(\Delta t) - U_{\rm BW}(\Delta t) || \leq \mathcal{E}(\Delta t)$, then it also follows that $1 - F_{\Delta t}^{\rm a} \leq \mathcal{E}(\Delta t)^{2}$, see App.~\ref{app: errors}.
For the local fidelity, we can calculate 
\begin{multline}
    \tilde{F}_{\Delta t}^{\rm a} = \frac{1}{N2^{N}} \sum_{\bm \sigma} \sum_{j=1}^{N} \langle {\bm \sigma} | U^{\dagger}_{\rm BW}(\Delta t) U_{\rm MPO}(\Delta t) \, \hat{\sigma}_{j} 
    \\
    U_{\rm MPO}^{\dagger}(\Delta t) U_{\rm BW}(\Delta t) | {\bm \sigma} \rangle .
    \label{fidelity_local_algorithmic}
\end{multline}
We explain how to calculate Eqs.~\eqref{fidelity_algorithmic} and \eqref{fidelity_local_algorithmic} using TNs in App.~\ref{app: errors}.
It is important to note that Eqs.~\eqref{fidelity_algorithmic} and \eqref{fidelity_local_algorithmic} do not give the infidelity (or a bound on the infidelity) for the evolution of arbitrary state $\ket{\psi}$.
Instead, they provide the infidelity averaged over all possible states.
Nevertheless, they will serve as a useful estimate of the algorithmic error for a single time step.

Equations \eqref{fidelity_algorithmic} and \eqref{fidelity_local_algorithmic} provide us with a way to estimate the fidelity of the time-evolved state with a BW circuit for the time step, $\Delta t$.
However, in practice, we would like to calculate the fidelity for $n$ applications of the circuit, i.e., the fidelity of evolving to time $t = n\Delta t$ using the product formula.
While calculating this exactly for large $N$ and large $t$ is generally not possible, we can again bound it by the operator norm error, $F^{\rm a}_{n\Delta t} \geq 1 - n^{2}\mathcal{E}(\Delta t)^{2}$.
We estimate that the total infidelity for $n$ applications of the circuit to go as
\beq
    I^{\rm a}_{n\Delta t} \approx n^{2}(1 - F_{\Delta t}^{\rm a}),
    \label{infidelity_multiple}
\eeq
and similarly for the local infidelity.
We expect that $1 - F^{\rm a}_{\Delta t} \sim \mathcal{O}(\Delta t^{2k+2})$ for a $k$th order approximation and small $\Delta t$, which is numerically verified for the CI model in App.~\ref{app: errors}.
Appendix~\ref{app: errors} also demonstrates that while \er{infidelity_multiple} can overestimate the true error for large $n$, it can still give a reasonable approximation.

\subsection{The total error}
We can now use our estimates of the fidelity to estimate a total fidelity for simulating a total time $t$ with $n$ time steps on a quantum device.
We expect the fidelity to be multiplicative,
\beq
    F_{n\Delta t} \approx e^{-\eta n}\left[1 - n^2(1 - F^{\rm a}_{\Delta t})\right],
    \label{total_error}
\eeq
where $\eta$ is the exponential decay rate of the fidelity due to native errors from the quantum device.
For some target simulation time, $t$, one can estimate the infidelity $I_{n\Delta t} = 1 - F_{n\Delta t}$ for a range of $n$, and use it to predict what choice of $\Delta t = t / n$ gives minimal infidelity.
While the values of $\eta$ recorded from the {\texttt ibm\_brisbane} are too large to reliably predict an optimal time, we show an example of calculating the optimal time step $\Delta t_{\rm opt}$ with fictional values of $\eta$ in App.~\ref{app: errors}.

\section{Conclusions} \label{sec: conclusions}
In this paper, we have proposed a method to classically optimize BW circuits to efficiently simulate the dynamics of quantum many-body systems,
building on the works of Refs.~\cite{Mansuroglu2023, Kotil2023, McKeever2023, Tepaske2023,Mansuroglu2023b}.
Like Refs.~\cite{McKeever2023, Tepaske2023}, we make use of MPO approximations of the time evolution operator to variationally update unitary gates in the BW circuits, allowing the method to efficiently scale with system size.
The previous works chose particular parameterisations of the two-qubit gates to enforce unitarity, which were optimised using gradient descent approaches. 
However, our method makes use of the polar decomposition to find efficient and optimal unitary gates \cite{Lin2021}.
We successively apply our method to multiple Hamiltonians with neighbouring three-body interactions (whereas previous works were limited to just two-body interactions), demonstrating that the dynamics of many-body Hamiltonians can be efficiently implemented using a BW structure with a shallow depth.
This is significant because BW circuits can easily be implemented on current digital quantum computers \cite{Lin2021}.

The first model we considered is a cluster Ising model.
We successfully demonstrated that dynamics of the model could be implemented with an accuracy that outperforms second order Trotterization with a BW circuit depth of $M = 3$, with significantly less CNOT layers required.
We then tested our method on the PXP model, a kinetically constrained model relevant to cold atom experiments. 
Here, we were also able to find circuits with depths $M = 4$ and $M = 5$ which outperform both first and second order Trotterization respectively.
Like the cluster Ising model, this is done with drastically less CNOT layers.
However, this came with the trade-off that the conserved quantities of the PXP model are violated (whereas Trotterization exactly conserves these quantities).
Finally, we tested our model on a next-nearing-neighbour Ising model.
Here, we were only able to find slight improvements on the accuracy when compared to second order Trotterization, but requires a small increase in the number of CNOT layers.
For this model, we found that our method often suffered from local minima; to overcome this, we used an initial ansatz inspired by the second order Trotterization.
This could motivate future studies to find strategies to overcome these optimisation difficulties. 

We have shown that our optimisation method can generate BW circuits which are more accurate and efficient than those produced by first and second order Trotterization, reducing the amount of algorithmic error.
This could allow us to simulate dynamics using larger time steps, and thus allow us to simulate dynamics for longer times before the onset of decoherence.
Furthermore, we show that the same circuits can be implemented on quantum computers using fewer layers of CNOTs than Trotterization.
This reduces the accumulated native errors of the quantum device, allowing us to simulate more time steps, again increasing the possible simulation window

We have also explained how the errors from optimisation and those from the quantum computer can be used to guide us in our choice of time step to minimise the total error.
We first implemented the optimised BW circuits and Trotterized circuits on IBM's quantum processor {\texttt ibm\_brisbane}, demonstrating that optimised BW circuits can achieve a more precise dynamics while being subject to less noise.
We then proposed that the optimal time step could be estimated through the infidelity of a single time step (with respect to the true time evolution operator), and the errors native to the quantum simulator.
This assumed that the algorithmic error from the optimisation of BW circuits grew linearly with the number of applications.
Indeed, for Trotterization it is known that this overestimates the error at large times \cite{Zhao2022}.
This estimation can be improved by finding tighter bounds on the errors \cite{Kivlichan2020, Childs2021, Layden2022}.
Similar advancements for BW circuits could facilitate a better choice of time step.
Alternatively, one could use adaptive methods to find the optimal choice of time step \cite{Zhao2023, Ikeda2023}.

One avenue for future works could be to investigate how to generalise the method to two dimensions.
While BW circuits are a natural ansatz for one-dimensional systems, two-dimensional systems can allow for more complex arrangements. 
This could allow for the possibility to also optimise the geometry of the circuits.
Furthermore, the use of TNs allowed our method to scale well with system size.
Unfortunately, two-dimensional TNs, such as projected-entangled pair states \cite{Verstraete2004b}, are known to suffer from a poor scaling with system size due to the inability to exactly contract them in an efficient manner \cite{Corboz2010, Lubasch2014}.
Nevertheless, there are promising developments in two-dimensional TNs which might allow for this to soon be a possibility \cite{Liao2019, Zaletel2020, Lin2022b, Geng2022}.
We hope to report on this at a later date.

\begin{acknowledgments}
    L.C. was supported by an EPSRC Doctoral prize from the University of Nottingham.
    A.M. acknowledges the support of the DAAD-WISE scholarship from the Deutscher Akademischer Austausch Dienst, and the KVPY fellowship from the Government of India. At the University of Toronto, A.M. is supported by the Lachlan Gilchrist Fellowship.
    F.P. acknowledges support from the Deutsche Forschungsgemeinschaft (DFG, German Research Foundation) under Germany’s Excellence Strategy--EXC--2111--390814868, the European Research Council (ERC) under the European Union’s Horizon 2020 research and innovation programme (Grant Agreement No. 851161), as well as the Munich Quantum Valley, which is supported by the Bavarian state government with funds from the Hightech Agenda Bayern Plus. 
    A.G.-S. acknowledges support from a research fellowship from the The Royal Commission for the Exhibition of 1851. 
    We acknowledge access to the University of Nottingham Augusta HPC service. 

\end{acknowledgments}

\section*{Code availability}
The code for optimising unitary brickwall circuits can be found at \href{https://zenodo.org/doi/10.5281/zenodo.10359963}{10.5281/zenodo.10359964}.

\appendix
\section{Error scaling in the trivial phase of the Cluster Ising model.} \label{app: errors_ci} 
\begin{figure}
    \centering
    \includegraphics[width=\linewidth]{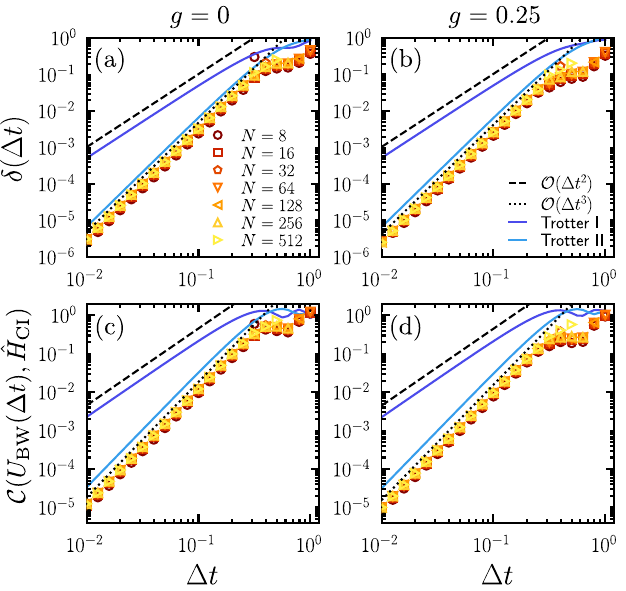}
    \caption{\textbf{Error scaling in the CI model.}
    (a, b) The error density, $\delta(\Delta t)$, and (c, d) the error in the conservation of energy, $\mathcal{C}(U_{\rm BW}(\Delta t), \hat{H}_{\rm CI})$ for the CI model with circuit depth $M = 3$.
    The left column is at the tricritical point $g = 0$, and the right column is in the trivial phase $g = 0.25$.
    All quantities are defined as in Fig.~\ref{fig: scaling_ci}.
    }
    \label{fig: ci_appendix}
\end{figure}

The variational optimisation method proposed in Sec.~\ref{sec: optimization} worked well for CI in the symmetry protected phase ($g < 0$) and had little dependence on initial state.
Conversely, the approach can struggle for $g \gtrsim 0$, often getting stuck in local minimum.
A reliable strategy to combat this is to anneal the parameter $g$ over the range $[-1, 1]$, using the result from a previous choice of $g$ as an initial seed.
In practice, we start from a value of $g = -1.0$ and optimise a random initial guess using our variational method.
We then iteratively reduce the parameter $g \leftarrow g - 0.05$ and optimise using the previous result as the initial guess. This is repeated until the target value of $g$ is obtained.
The results of this approach are shown in Fig.~\ref{fig: ci_appendix} for $g = 0$ and $g = 0.25$ with circuit depth $M = 3$.
The top panel shows the error density and the bottom panel shows the error in the conservation of energy.
The results closely match that of Fig.~\ref{fig: scaling_ci}.

\section{Implementing brickwall and Trotterized circuits on quantum computers} \label{app: decompositions} 
\begin{figure*}
    \centering
    \includegraphics[width=0.8\linewidth]{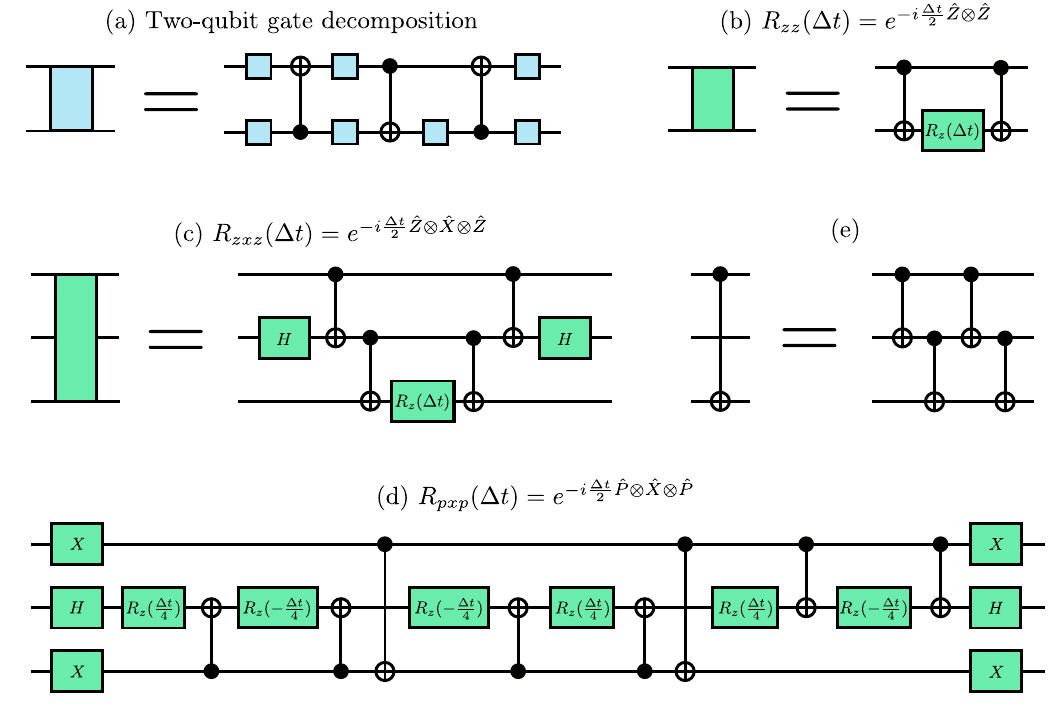}
    \caption{\textbf{Decompositions of unitary gates.}
    (a) A decomposition for any two-qubit unitary gate in terms of single-qubit unitaries and three CNOT gates.
    (b) The two-qubit rotation gate $R_{zz}(\Delta t) = e^{-i\frac{\Delta t}{2}\hat{Z}\otimes\hat{Z}}$ can be implemented using only two CNOT gates, and a single-qubit rotation $R_{z}(\Delta t) = e^{-i\frac{\Delta t}{2}\hat{Z}}$.
    (c) The three-qubit gate $R_{zxz}(\Delta t) = e^{-i\frac{\Delta t}{2}\hat{Z}\otimes\hat{X}\otimes\hat{Z}}$ can be decomposed as four CNOT gates, an $R_{z}(\Delta t)$ gate and two Hadamard gates, denoted by $H$.
    (d) The PXP gate, $R_{pxp}(\Delta t) = e^{-i\frac{\Delta t}{2} \hat{P}\otimes\hat{X}\otimes\hat{P}}$, can be decomposed into eight CNOT gates, six $R_{z}(\pm \Delta t/4)$ rotation gates, two Hadamard gates and four Pauli-X gates, denoted by $X$.
    (e) A CNOT gate which acts on two next-nearest-neighbouring qubits can be decomposed into 4 CNOTs which act only on neighbouring qubits.}
    \label{fig: decompositions}
\end{figure*}

The first unitary gate we consider is a general two-qubit gate, such as those in unitary BW circuits.
Any two-qubit gate can be decomposed into a circuit of CNOT gates and single-qubit gates, which requires at most three CNOTs.
This is shown in Fig.~\ref{fig: decompositions}(a).
The single-qubit gates need to be carefully chosen to make this decomposition valid; see Refs.~\cite{Kraus2001, Smith2019} for more details.
While the general decomposition can be achieved with three CNOTs, some gates can be implemented with less.
For example, the two-qubit rotation gate 
\beq
    R_{zz}(\Delta t) = e^{-i\frac{\Delta t}{2}\hat{Z}\otimes\hat{Z}}
\eeq
can be implemented with a single-qubit rotation gate $R_{z}(\Delta t) = e^{-i\frac{\Delta t}{2} \hat{Z}}$ placed on the target qubit in-between two identical CNOTs \cite{Nielsen2011}.
Notice that this gate is used in the Trotterization for both the CI model and the NNNI model.

Three-qubit entangling gates require more care to implement.
For rotations of Pauli strings on three qubits, it is well understood how to efficiently implement such gates using only single-qubit gates and CNOTs \cite{Whitfield2011}.
For example, the rotation gate
\beq
    R_{zxz}(\Delta t) = e^{-i\frac{\Delta t}{2}\hat{Z}\otimes\hat{X}\otimes\hat{Z}},
\eeq
is shown in Fig.~\ref{fig: decompositions}(c).
We first note that
\beq
    R_{zxz}(\Delta t) = (\hat{\mathds{1}} \otimes H \otimes \hat{\mathds{1}}) \, R_{zzz}(\Delta t) \, (\hat{\mathds{1}} \otimes H \otimes \hat{\mathds{1}}),
\eeq
with $R_{zzz}(\Delta t) = e^{-i\frac{\Delta t}{2}\hat{Z}\otimes\hat{Z}\otimes\hat{Z}}$.
The $R_{zzz}(\Delta t)$ gate can then be decomposed into four CNOTs and an $R_{z}(\Delta t)$ gate \cite{Nielsen2011}.

To implement the three-body gates in the Trotterized circuits for the PXP model, we first note that
\begin{multline}
    R_{pxp}(\Delta t) = e^{-i\frac{\Delta t}{2}\hat{P}\otimes\hat{X}\otimes\hat{P}} 
    \\
    = (\hat{X} \otimes H \otimes \hat{X}) \, R_{qzq}(\Delta t) \, (\hat{X} \otimes H \otimes \hat{X}),
\end{multline}
where $\hat{Q} = \hat{\mathds{1}} - \hat{P}$ and $R_{qzq}(\Delta t) = e^{-i\frac{\Delta t}{2}\hat{Q}\otimes\hat{Z}\otimes\hat{Q}}$.
We can then write
\begin{multline}
    R_{qzq}(\Delta t) = (\hat{Q} \otimes R_{z}(\Delta t) \otimes \hat{Q}) + (\hat{P} \otimes \hat{\mathds{1}} \otimes \hat{Q})
    \\
    + (\hat{Q} \otimes \hat{\mathds{1}} \otimes \hat{P}) + (\hat{P} \otimes \hat{\mathds{1}} \otimes \hat{P}).
\end{multline}
This can be considered to be a three-qubit gate where the first and last qubit act as control gates for the rotation gate on the central qubit.
The remainder of the circuit makes use of six $R_{z}(\Delta t)$ gates and eight CNOT gates to implement the Trotterized PXP dynamics \cite{Nielsen2011}, and is shown in Fig.~\ref{fig: decompositions}(d).
While we will not give the complete details of how the circuit is derived, one can carefully check that this correctly implements the gate $R_{pxp}(\Delta t)$.
Notice that the circuit has two CNOT gates which act on next-nearest-neighbouring qubits.
If the quantum computer simulating the dynamics has connections between these qubits, then the circuit can be used as presented.
However, if the quantum computer only has connections between neighbouring qubits, then these CNOT gates must be decomposed into CNOTs acting only on neighbouring qubits; the most optimal known decomposition for this is shown in Fig.~\ref{fig: decompositions}(e), and requires four CNOTs \cite{Shende2006}.
This increases the number of CNOTs required for the $R_{pxp}(\Delta t)$ gate to fourteen.

\subsection{The number of CNOT layers required for brickwall circuits and Trotterization}
We can now calculate the number of CNOT layers required to implement both the BW circuits and Trotterized circuits for each of the models considered in Sec.~\ref{sec: results}.
BW circuits require three CNOT layers per BW layer.
For the CI model, a BW circuit with depths $M = 3$ and $M = 4$ both give a second order scaling, requiring $9$ and $12$ CNOT layers. respectively.
First order Trotterization for the CI model requires two layers of $R_{zz}(\Delta t)$ gates and three layers of $R_{zzz}(\Delta t)$ gates, with each layer requiring two and four CNOTs respectively.
This gives a total of $16$ CNOT layers for the whole circuit.
Second order Trotterization requires four layers of $R_{zz}(\Delta t)$ gates and five layers of $R_{zzz}(\Delta t)$ gates, giving a total of $28$ CNOT layers.

For the PXP model, first and second order scaling are achieved with circuit depths $M = 4$ and $M = 5$ respectively for the BW circuit. 
These require a total of $12$ and $15$ CNOT layers.
First order Trotterization requires three layers of $R_{pxp}(\Delta t)$ gates, each of which can be implemented using fourteen layers of {\em neighbouring} CNOTs, giving a total of $42$ CNOT layers.
Similarly, second order Trotterization requires five layers of $R_{pxp}(\Delta t)$ gates, giving a total of $70$ CNOT layers.
If next-nearest-neighbours connections are available, then this can be reduced to $24$ and $40$ CNOT layers respectively.

BW circuits with depths $M = 5$ or $M = 6$ are required to yield a second order scaling for the NNNI model, which can be implemented with a total of $15$ and $18$ CNOT layers respectively.
The optimal second order Trotterization is shown in Fig.~\ref{fig: scaling_nnni}(a).
Note that while the structure of the circuit looks similar to that of the BW circuit, the two-qubit entangling gates here are exclusively SWAP gates (which require three CNOTs) and $R_{zz}(\Delta t)$ gates.
$R_{zz}(\Delta t)$ can be implemented using only two CNOTs, compared to the more general two-qubit gate which requires three CNOTs.
This gives an improvement, only requiring a total of $13$ CNOT layers if $g_{zz} = 0$, or $15$ CNOT layers if $g_{zz} \neq 0$.

\section{Native and algorithmic errors for simulating dynamics} \label{app: errors} 
In this Appendix, we provide further details for the discussion in Sec.~\ref{sec: ibm}.
We first prove that the infidelity of a unitary approximation is bounded by its operator norm error.
We then show how the infidelity averaged over all initial basis states can be calculated using TNs, and how this error propagates after multiple applications of the unitary approximation.
Finally, we use our results to determine optimal time steps for simulating dynamics.

\subsection{Error bounds on the infidelity}
Given the exact unitary operator, $U(\Delta t) = e^{-i\Delta t\hat{H}}$, and some unitary approximation, $V(\Delta t)$, we want to find an upper bound on the infidelity 
\beq
    I_{\psi} = 1 - |\braket{\psi | U^{\dagger}(\Delta t)^{n} V(\Delta t)^{n} | \psi}|^{2}
\eeq
for some normalised wavefunction $\ket{\psi}$.
We first assume that the error of the approximation is $||U(\Delta t) - V(\Delta t)|| = \mathcal{E}(\Delta t)$, where $\| \cdot \|$ is the operator norm of a matrix.
This implies that $\|[U(\Delta t) - V(\Delta t)] \ket{\psi}\| \leq \mathcal{E}(\Delta t) \|\ket{\psi}\|$ for all $\ket{\psi}$, where $\|\ket{\psi}\| = \braket{\psi | \psi}$ is the Euclidean vector norm.
We can then bound the error for the product formula,
\begin{widetext}
\begin{multline}
    \| U(\Delta t)^{n} - V(\Delta t)^{n} \| = \bigg|\bigg| \sum_{m=0}^{n-1} U(\Delta t)^{m} [U(\Delta t) - V(\Delta t)] V(\Delta t)^{n-m-1}\bigg|\bigg|
    \leq \sum_{m=0}^{n-1} \big\| U(\Delta t)^{m}[U(\Delta t) - V(\Delta t)] V(\Delta t)^{n-m-1} \big\|
    \\
    \leq \sum_{m=0}^{n-1} \| U(\Delta t)^{m}|| \, \| U(\Delta t) - V(\Delta t) \| \, \| V(\Delta t)^{n-m-1} \|
    = \sum_{m=0}^{n-1} \| U(\Delta t) - V(\Delta t) \| = n\mathcal{E}(\Delta t).
\end{multline}
\end{widetext}
The first equality can be shown by induction, and the inequalities follow from the properties of the operator norm.
It then follows that 
\begin{multline}
    {\rm Re} \left(\braket{\psi | U^{\dagger}(\Delta t)^{n} V(\Delta t)^{n} | \psi}\right) 
    \\
    = 1 - \frac{1}{2} \big|\big|[U(\Delta t)^{n} - V(\Delta t)^{n}] \ket{\psi}\big|\big|^{2} 
    \\
    \geq 1 - \frac{1}{2}n^{2}\mathcal{E}(\Delta t)^{2}.
\end{multline} 

\noindent This allows us to bound the infidelity by the error of the unitary approximation,
\beq
    I_{\psi} \leq 1 - \left(1 - \frac{1}{2}n^{2}\mathcal{E}(\Delta t)^2\right)^2 \leq n^{2}\mathcal{E}(\Delta t)^{2}.
    \label{infidelity_bound}
\eeq
Notice that this holds for all normalised $\ket{\psi}$.

\subsection{Estimating the state-averaged infidelity using TNs}

\begin{figure}[h]
    \centering
    \includegraphics[width=\linewidth]{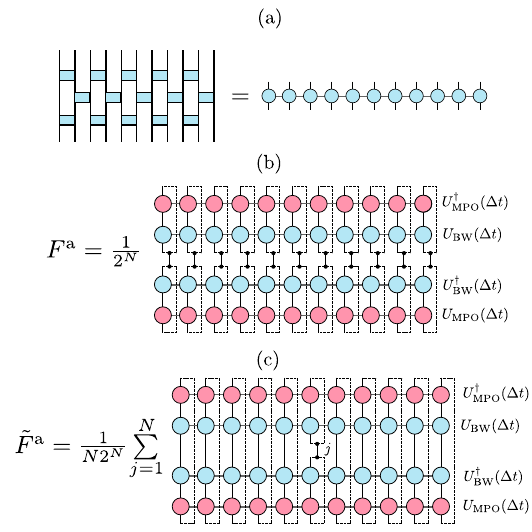}
    \caption{\textbf{Fidelity calculations as TNs.}
    (a) A BW circuit with circuit depth $M$ can be written as an MPO with bond dimension $\chi = 2^{\lceil M/2 \rceil}$.
    (b) The fidelity \er{fidelity_algorithmic} can be calculated as a TN; the (almost) exact time propagator, and its adjoint, are represented by an MPO with the pink circles.
    The BW circuit is an MPO with blue circles.
    The black dots are three-point delta functions.
    (b) The local fidelity at site $j$ is calculated in the same way, but the delta functions are only placed at site $j$ (with a sum over $j$).
    }
    \label{fig: fidelity_calc}
\end{figure}

The algorithmic fidelity \er{fidelity_algorithmic} and the local fidelity \er{fidelity_local_algorithmic} can be well estimated using the MPO approximation $U(\Delta t) \approx U_{\rm MPO}(\Delta t)$ and the BW circuit $U_{\rm BW}(\Delta t)$.
It is first convenient to write the BW circuit as an MPO, which can be achieved exactly using bond dimension $\chi = 2^{\lceil M/2 \rceil}$ (although in practice it might be advantageous to compress this into an MPO with smaller bond dimension).
This is shown in Fig.~\ref{fig: fidelity_calc}(a).
To calculate the fidelity, we can write
\begin{widetext}
\begin{multline}
    F_{\Delta t}^{\rm a} = \frac{1}{2^{N}} \sum_{\sigma_{1}\cdots \sigma_{N}} \sum_{\sigma'_{1}\cdots \sigma'_{N}} \braket{\sigma_{1}\cdots \sigma_{N} | U_{\rm MPO}^{\dagger}(\Delta t) U_{\rm BW}(\Delta t) | \sigma'_{1}\cdots \sigma'_{N}}
    \\
    \sum_{\phi_{1}\cdots \phi_{N}} \sum_{\phi'_{1}\cdots \phi'_{N}} \braket{\phi_{1}\cdots \phi_{N} | U_{\rm BW}^{\dagger}(\Delta t) U_{\rm MPO}(\Delta t) | \phi'_{1}\cdots \phi'_{N}}
    \prod_{j=1}^{N} \sum_{k_j} \delta_{\sigma_{j}, \sigma'_{j}, k_{j}} \delta_{\phi_{j}, \phi'_{j}, k_{j}}
    \label{fidelity_algorithmic_long}
\end{multline}
\end{widetext}
where $\delta_{\sigma_{j}, \sigma'_{j}, k_{j}}$ is the three-point delta functions for spins $\sigma_{j}$, $\sigma'_{j}$ and $k_{j}$.
The delta functions enforce the condition $\sigma_{j} = \sigma'_{j} = \phi_{j} = \phi'_{j}$.
Despite the fact that \er{fidelity_algorithmic_long} appears more complicated than \er{fidelity_algorithmic}, it is very convenient to write the fidelity in this way if we want to calculate it as a TN.
The TN representation is shown in Fig.~\ref{fig: fidelity_calc}(b), where the black dots represent the three-point delta functions.

The algorithmic local fidelity can be calculated similarly.
Here, we write 
\begin{widetext}
    \begin{multline}
        \tilde{F}_{\Delta t}^{\rm a} = \frac{1}{N2^{N}}\sum_{j=1}^{N} \sum_{\sigma_{1} \cdots \sigma_{N}} \sum_{\sigma'_{1} \cdots \sigma'_{N}} \braket{\sigma_{1}\cdots \sigma_{j}\cdots\sigma_{N} | U_{\rm MPO}^{\dagger}(\Delta t) U_{\rm BW}(\Delta t) | \sigma'_{1}\cdots \sigma'_{j}\cdots\sigma'_{N}}
        \\
        \sum_{\phi_{j}, \phi'_{j}}  \braket{\sigma'_{1}\cdots \phi_{j} \cdots \sigma'_{N} | U_{\rm BW}^{\dagger}(\Delta t) U_{\rm MPO}(\Delta t) | \sigma_{1}\cdots \phi'_{j} \cdots \sigma_{N}}
        \sum_{k_j} \delta_{\sigma_{j}, \sigma'_{j}, k_{j}} \delta_{\phi_{j}, \phi'_{j}, k_{j}}.
        \label{local_fidelity_algorithmic_long}
    \end{multline}
\end{widetext}
Notice that we only need to introduce the delta functions at spin $j$.
This is shown as a TN in Fig.~\ref{fig: fidelity_calc}(c).

\begin{figure}[h]
    \centering
    \includegraphics[width=\linewidth]{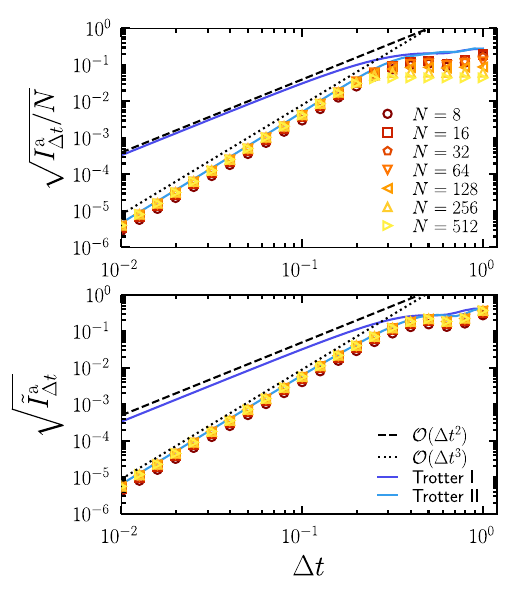}
    \caption{\textbf{The algorithmic infidelity.}
    The square root of the algorithmic infidelity scaled by system size, $\sqrt{I^{\rm a}_{\Delta t} / N}$, (top panel) and the square root of the algorithmic local infidelity, $\sqrt{\tilde{I}^{\rm a}_{\Delta t}}$ (bottom panel), as functions of the time step, $\Delta t$.
    Results are for the CI model with $g = -0.75$ and circuit depth $M = 3$.
    The data points and curves are as defined in Fig.~\ref{fig: scaling_ci}.
    }
    \label{fig: fidelity_scaling}
\end{figure}

Figure~\ref{fig: fidelity_scaling} shows the square root of infidelity and local infidelity for optimised BW circuits with depth $M = 3$, calculated using Eqs.~\eqref{fidelity_algorithmic_long} and \eqref{local_fidelity_algorithmic_long}.
The plot for the fidelity is scaled by system size.
The infidelities are obtained using $I_{\Delta t}^{\rm a} = 1 - F_{\Delta t}^{\rm a}$ and $\tilde{I}_{\Delta t}^{\rm a} = 1 - \tilde{F}_{\Delta t}^{\rm a}$.
As was the case for the error density, both of these quantities scale as $\mathcal{O}(\Delta t^{3})$ respectively (dotted line), and outperform the results of both first and second order Trotterization (solid lines).

\subsection{State-averaged infidelity of the product formula}

We now comment on the propagation of algorithmic errors when using the product formula $U(t = n\Delta t) \approx U_{\rm BW}(\Delta t)^{n}$.
In the main text, the estimation of the infidelity $I^{\rm a}_{n\Delta t} \approx n^{2} I^{\rm a}_{\Delta t}$ was motivated by the bound from the error.
For small system sizes, we are able to calculate both $U(n\Delta t)$ and $U_{\rm BW}(\Delta t)^{n}$ exactly and decompose them as MPOs.
This allows us to use Eqs.~\eqref{fidelity_algorithmic_long} and \eqref{local_fidelity_algorithmic_long} and thus calculate the infidelity using the product formulae.

\begin{figure}
    \centering
    \includegraphics[width=\linewidth]{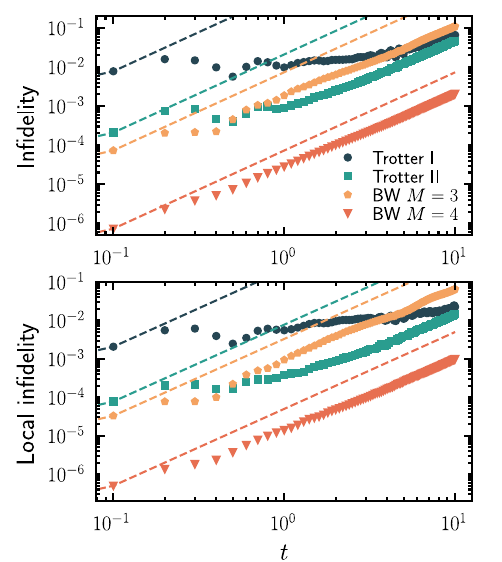}
    \caption{\textbf{The propagated algorithmic infidelity.}
    The (top) infidelity and (bottom) local infidelity of $U_{\rm BW}(\Delta t)^{n}$ when compared to the exact time evolution operator  $U_{\rm BW}(n\Delta t)$ as a function of $t = n\Delta t$.
    The results are shown for the CI model with $g = -0.75$ and $N = 8$, with time step $\Delta t = 0.1$.
    Results for both BW circuits and Trotterization are shown.
    The dashed lines show the estimate $I^{\rm a}_{n\Delta t} \approx n^{2} I^{\rm a}_{\Delta t}$.
    }
    \label{fig: infidelity_product}
\end{figure}

Figure~\ref{fig: infidelity_product} shows the results for the CI model with $g = -0.75$ and $N = 8$ with time step $\Delta t = 0.1$.
We show both BW circuits with depths $M = 3$ and $M = 4$, and also first and second order Trotterization.
The dashed lines show the estimate.
Notice that at early times, both BW circuit depths outperform both orders of Trotterization.
However, this is superseded at later times, when both first and second order Trotterization perform better than the BW with depth $M = 3$.
While this is surprising, it is consistent with recent results \cite{Layden2022} which have shown that the time-propagated Trotter error can have stricter bounds than originally expected \cite{Lloyd1996}, albeit for systems with nearest-neighbour interactions.
It is important to note that, however, the BW circuit with depth $M = 3$ is still subject to significantly less noise than both first and second order Trotterization due to its shallower quantum circuit implementation. 
The same is true for depth $M = 4$, which has a smaller time-propagated infidelity than both orders of Trotterization.
In each case, the estimate over-estimates the true error.
Nevertheless, it serves as a useful estimate for both BW circuits.

\subsection{Predicting an optimal time step}
\begin{figure}[t]
    \centering
    \includegraphics[width=\linewidth]{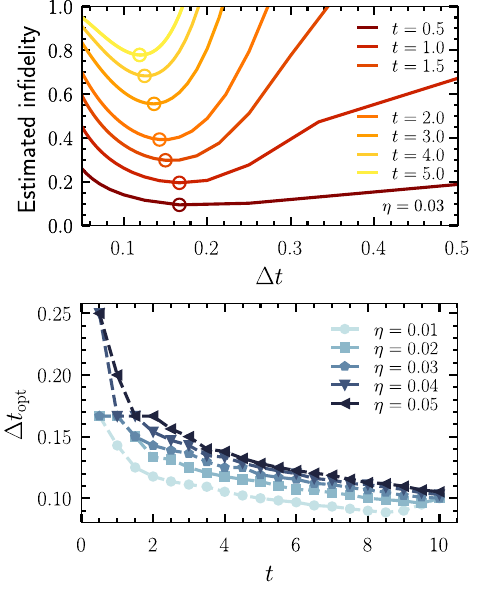}
    \caption{\textbf{Predicting an optimal time step.}
    The algorithmic infidelity and the error from the device can be used to predicted an optimal time step.
    (a) The estimated infidelity using \er{total_error} as a function of time step, $\Delta t$, for the CI model with $g = -0.5$, $N = 5$ and $M = 3$. 
    Each curve shows the estimated  different total simulated time, $t$.
    The markers indicate the optimal time step, $\Delta t_{\rm opt}$.
    The curves are shown for fictional $\eta = 0.03$.
    (b) The optimal time step, $\Delta t_{\rm opt}$, as a function of $t$ for various fictional $\eta$.
    }
    \label{fig: infidelity_predicted}
\end{figure}

We now demonstrate how the algorithmic infidelity and the infidelity measured from the quantum device can be used to predict an optimal time step to minimise the total error.
Note that, unfortunately, the errors measured from the {\texttt ibm\_brisbane} device are too large to do our analysis.
We instead use \er{total_error} with some fictional choices of $\eta$.

Figure~\ref{fig: infidelity_predicted}(a) shows \er{total_error} as a function of time step, $\Delta t$, for various total simulation times, $t$, and $\eta = 0.03$.
Notice that for small time steps, $\Delta t$, the estimated infidelity is large due to the number of applications of BW circuit that is needed.
On the contrary, for large $\Delta t$, the error is large because the BW approximation has substantial errors.
The optimal time steps, shown by the circle markers, is the choice which balances the two errors.
This is shown in Fig.~\ref{fig: infidelity_predicted}(b) as a function of $t$ for multiple $\eta$.
Notice that the optimal time step decreases with $t$.
While we have also observed this is the case for other choices of $g$, the exact details might depend on the problem.

%\bibliography{bibliography}
\bibliographystyle{apsrev4-2}

\end{document}